\documentclass[preprint2]{aastex}
\usepackage{graphicx}
\usepackage{amsmath}
\usepackage{txfonts}
\usepackage{rotating}
\usepackage{lscape}
\usepackage{textcomp}
\usepackage{amssymb}

\begin{document}

\shorttitle{A new extensive catalog of optically variable AGN in the GOODS
Fields}
\shortauthors{Villforth et al.}

\title{A new extensive catalog of optically variable AGN in the GOODS Fields
and a new statistical approach to variability selection \footnote{Based on
observations obtained with the NASA/ESA Hubble Space Telescope, which is operated
by the Association of Universities for Research in Astronomy (AURA), Inc., under
NASA contract NAS5-26555.} } \author{ Carolin Villforth\altaffilmark{1,2}, Anton
M. Koekemoer\altaffilmark{1} and Norman A. Grogin\altaffilmark{1}}

\altaffiltext{1}{Space Telescope Science Institute, 3700 San Martin Drive,
Baltimore, MD 21218, USA}
\altaffiltext{2}{University of Turku, Department of Physics and Astronomy,
Tuorla Observatory, V\"{a}is\"{a}l\"{a}ntie 20, 21500 Piikki\"{o}, Finland}

\begin{abstract}
Variability is a property shared by practically all AGN. This makes variability
selection a possible technique for identifying AGN. Given that variability
selection makes no prior assumption about spectral properties, it is a powerful
technique for detecting both low-luminosity AGN in which the host galaxy emission
is dominating and AGN with unusual spectral properties. In this paper, we will
discuss and test different statistical methods for the detection of variability
in sparsely sampled data that allow full control over the false positive rates.
We will apply these methods to the GOODS North and South fields and present a
catalog of variable sources in the $z$ band in both GOODS fields.
Out of 11931 objects checked, we find 155 variable sources at a
significance level of 99.9\%, corresponding to about 1.3\% of all objects. After
rejection of stars and supernovae, 139 variability selected AGN remain. Their magnitudes reach down as faint as 25.5 mag
in $z$. Spectroscopic redshifts are available for 22 of the variability selected
AGN, ranging from 0.046 to 3.7. The absolute magnitudes in the rest-frame z-band
range from $\sim$ -18 to -24, reaching substantially fainter than the typical
luminosities probed by traditional X-ray and spectroscopic AGN selection in these
fields. Therefore, this is a powerful technique for future exploration of the
evolution of the faint end of the AGN luminosity function up to high redshifts.
\end{abstract}


\section{Introduction}

In the last 20 years or so, findings of strong correlations between the black
hole masses in the centers of nearby galaxies and the properties of their hosting
galaxies
\citep[e.g.][]{ferrarese_fundamental_2000,graham_correlation_2001,marconi_relation_2003,gebhardt_relationship_2000}
have moved Active Galactic Nuclei (AGN) into the center of attention as
key players in galaxy evolution.

However, while it is clearly established that AGN must play a major
role in galaxy formation to be able to produce the relation between supermassive
black holes and their host galaxies at redshift zero, it is still a  mystery
when the correlation came into place and through which mechanism. To understand
this, one has to understand both the evolution of AGN and their connection with their
hosting galaxies.

While our knowledge of low-redshift AGN and their correlation with the host is
relatively broad, much less is known about high-redshift AGN hosts. Studies of
AGN hosting galaxies seldom reach beyond redshifts of two. Resolving the host
galaxy becomes more and more challenging as the faint hosts are extremely hard to
disentangle from the nuclear emission
\citep[e.g.][]{hutchings_host_2003,jahnke_ultraviolet_2004,schramm_host_2008,villforth_quasar_2008}.
This is due to both the decrease in apparent host galaxy size and surface
brightness dimming. Low-luminosity AGN have a much more favorable core-to-galaxy
ratio, making it possible to study the properties of their hosting galaxies up to
very high redshifts.

Another open question is the general
evolution of AGN over redshift. It is known that the luminosity function of
high-redshift AGN significantly differs from the one at low redshifts.
High-redshift AGN are on average much brighter than their low-redshift counterparts \citep{dunlop_redshift_1990}.
However, given the fact that almost nothing is known about low-luminosity AGN at
high-redshift, this is just the tip of the iceberg and we know little about the
shape and normalization of the luminosity function on the faint end.

New studies also imply that high-redshift AGN might be intrinsically different
from their low-redshift counterparts. \cite{jiang_dust-free_2010} found an
interesting sample of dust-free AGN at high-redshift that have no low-redshift
counterparts. And \cite{shemmer_x-ray_2009} found a sample of weak-lined
high-redshift AGN that do not seem to have  low-redshifts counterparts. On the
other hand, the interesting AGN subclass of BL Lacs, highly variable objects
thought to be the beamed counterparts of Fanarof-Riley I radio galaxies
\citep{urry_unified_1995} so far have only been detected at low-redshift. The
missing high-redshift BL Lacs are still a mystery given their extreme intrinsic
brightness \citep{stocke_hidden_2001}. Latest theoretical models for the
evolution of AGN imply that the missing BL Lac problem might be due to the fact that
certain types of AGN preferably appear at certain redshifts, with Fanarof-Riley
type I galaxies being extremely rare at high redshifts
\citep{garofalo_evolution_2010}. Studying low-luminosity AGN at high-redshift
might therefore also help answering the question what determines the intrinsic
properties of AGN.

It is also still a puzzle how supermassive black holes formed. The fact that AGN
at redshifts greater than six are found to host supermassive black holes as
large as $10^{9} M_{\odot}$ \citep{jiang_dust-free_2010} is rather puzzling
given the fact that the universe was less than a gigayear old at those
redshifts. This poses the question of how and when those black holes
were formed. Getting a more complete view of black hole masses at high redshift might help understand
how, when and in which objects mass accumulation took place.

Despite the great interest in this topic, our knowledge about high-redshift AGN
is still limited and almost nothing is known about faint AGN at high redshifts.
To learn more about this topic, we need new complete samples of high-redshift AGN
reaching to much lower luminosities. Deep multi-band surveys offer the
possibility to select such samples.

AGN can be identified in several ways, the most obvious being through
spectroscopy. AGN can either be identified through extremely broad lines or by
determining line ratios
\citep{baldwin_classification_1981,veilleux_spectral_1987}. The former has the
problem that it only selects the small subset of broad-lined Type I again and
misses Type II AGN as well as weak-lined objects. Both have the problem that
spectroscopy is extremely costly especially for faint sources. Therefore, while
spectroscopic surveys can play an important role in understanding high redshift
AGN, carefully selected candidate samples are necessary to make this method
efficient.

A commonly used method for AGN selection is through their optical colors. This
method was first used by \cite{markarian_galaxies_1967} who selected objects with
excess UV flux and created the first catalog of nearby AGN. Possibly the most
famous color selected AGN catalog is the Palomar Green Bright Quasar Catalog
\citep[][known under the name BQS or PG]{schmidt_quasar_1983} in which objects
with $U-B < -0.44$ mag were selected. However, UV excess might also be caused by
star formation. Therefore more sophisticated methods are needed to avoid
contamination by star forming galaxies. \cite{warren_quasar_1991} designed a
multi-color selection method, in which only point-like sources are chosen and
stars are rejected by their specific location in a multi-dimensional color
space. Similar methods have been used for quasar selection in the Sloan Digital
Sky Survey (SDSS) \citep{richards_spectroscopic_2002}. Such methods
are observationally cheap and can easily be applied to big multi-waveband surveys. However, they rely on considerable deviation from a normal stellar-dominated
spectral energy distribution (SED). Therefore such methods are not suitable to detect the
interesting sample of low-luminosity AGN in which the galaxy emission dominates
the overall SED.

Other ways to select AGN are through excess radio, X-ray or mid-infrared
emission. Radio selection has the downside that it only selects a small subsample
of AGN. Only about 10\% of all optically-selected AGN show considerable radio
emission \citep{smith_radio_1980}. Radio surveys are also generally shallow or
may suffer from confusion due to large apertures. An expection to this
rule is the FIRST Survey which imaged a large area of the sky using the NRAO
Very Large Arry, the resolution of this survey is similar to ground-based optical surveys
and the depth of the survey exceeds most other radio surveys
\citep{becker_first_1995}. X-ray surveys have similar problems to most radio
surveys, the resolution is generally rather poor, causing problems with
confusion. Due to the limited size of space-based X-ray telescopes it is also hard to detect faint high redshift sources (Chandra has a 1.2 m mirror \citep{weisskopf_chandra_2000}, the XMM-Newton effective collecting area corresponds to a mirror size of only about 80cm \citep{jansen_xmm-newton_2001}).

Mid-infrared emission is thought to originate from warm dust in the obscuring
dust torus surrounding the AGN \citep{sanders_continuum_1989}. However, also
star forming galaxies such as ultra luminous infrared galaxies (ULIRGs) show
excess mid-infrared emission \citep{genzel_what_1998}. As star forming galaxies
are especially common at redshifts around 2 or 3 \citep[see
e.g.][]{madau_star_1998,reddy_spectroscopic_2006,bouwens_discovery_2010},
mid-infrared selected samples are polluted by star forming galaxies, especially
at higher redshifts. Additionally, at very high-redshifts, the Universe might
have been too young for considerable amounts of dust to be produced. Indeed,
some high-redshift AGN have been shown to be virtually dust-free
\citep{jiang_dust-free_2010}.

Given that practically all AGN vary on all timescales from hours to decades
\citep[for a review, see][]{ulrich_variability_1997}, variability can be used as
a selection criterion for AGN \citep[e.g.][]{sarajedini_v-band_2003}. Due to
light time travel arguments, any variability detected in galaxies on human-observable
time-scales must originate from the nuclear region. Interestingly, it has also
been found that fainter AGN vary more strongly than their bright counterparts
\citep[e.g.][]{trevese_ensemble_1994,cristiani_optical_1996,
di_clemente_variability_1996,de_vries_structure_2005,wold_dependence_2007}. This
makes variability-selected samples especially sensitive to the interesting and
otherwise difficult to detect sample of low-luminosity AGN. Additionally, it has
been found that AGN vary more strongly on shorter wavelength
\citep[e.g.][]{di_clemente_variability_1996}. This makes variability selection
more sensitive at higher redshift in a given optical waveband.

Deep multi-waveband surveys such as the Great Observatories Origins Deep Survey
(GOODS) \citep{giavalisco_great_2004} are very well suited for creating
such a variability-selected AGN sample. Data for such surveys are typically taken in
several epochs, distributed over several months and are therefore suitable to
detect variabitility on timescales of months. Additionally, extremely
deep imaging on a wide range of wavelengths all the way from X-ray to radio gives the possibility to study
the broadband multiwavelength properties of variability selected AGN
\cite[e.g.][]{paolillo_prevalence_2004} as well as their parent population.

\cite{sarajedini_v-band_2003} were the first to attempt assembling a variability
selected sample from multi-epoch survey data in the GOODS Fields. They
used the two-epoch Hubble Deep Field (HDF) $V$-band data, sampling a time-span of 5 years. They found evidence
for nuclear variability in 16 galaxies down to a $V$-band magnitude of 27.5.
Given that this study was performed on only two epochs of data, this was an
extremely encouraging result, showing that variability selection can succeed even
for sparsely sampled data. The AGN found by \cite{sarajedini_v-band_2003} show
redshifts up to 1.8. They were able to show that the variability
selected AGN cover a wide range of colors and might not have been detected using color-color selection criteria.
Based on their sample they derived an AGN luminosity function and found that
low-luminosity AGN are possibly more abundant at high redshifts.

\cite{cohen_clues_2006} published a similar study based on the Hubble Ultra Deep
Field (HUDF) $i$-band data, sampling a time-span of about 4 months. They found
about 1\% of the sources to show significant variability with
photometric redshifts as high as 5.

\cite{klesman_optical_2007} presented a study of the five epoch $V$-band data in
the GOODS South field. They limited their study to infrared power-law and X-ray
selected sources. As much as 26\% of their AGN candidate sample showed
variability.

\cite{trevese_variability-selected_2008} applied similar method to the Southern
inTermediate Redshift ESO Supernova Search (STRESS) data. This survey
covers about five square degrees around the Chandra Deep Field South. In
contrast to previous studies discussed here, the data were taken with a
ground-based telescope. This results in significantly lower resolution, making it
harder to detect the variability of a point source against the flux of its
hosting galaxy. They applied their method to eight epochs of $V$ band data taken
around the Chandra Deep Field South (CDFS) and found 112 out of 5138
objects (about 2\% of all sources) to be variable.

Detecting variability from sparsely sampled data is extremely challenging. In
general, there are two approaches to variability detection. The first approach is
to use well calibrated and robust statistical estimators such as the
$\chi^{2}$ statistics. This approach allows for an effective control of false
positive rates as the estimators have known expected distributions for samples of
non-variable sources. The downside of such an
approach is that one has to rely on the correctness of the error measurements.
Given the fact that all kinds of unknown sources of error and systematics might influence
aperture photometry, this approach is often avoided. The only statistical
estimator for variability selection that does not show this drawback is the
ANalysis Of VAriance (ANOVA) \citep{de_diego_testing_2010}. It can
however only be applied to data with large numbers of data points that is oversampled in the time domain \citep[see
e.g.][]{villforth_intranight_2009,de_diego_testing_2010}.

The second approach derives 'typical' scatter at a certain magnitude and then
declares a certain number of objects that show 'significantly more scatter' 
variable \cite[this approach has been taken
by][]{sarajedini_v-band_2003,cohen_clues_2006,klesman_optical_2007,trevese_variability-selected_2008}.
While this approach is robust in a sense that the scatter in the 'real' data is
taken into account, it is lacking a control of false positives and locally
differing errors.

In this paper, we will present a method that combines the advantages of both
approaches. Our methodology takes into account the observed scatter but uses
well-calibrated statistical estimators. We are therefore in full control of
expected false positive rates.

We present a catalog of 139 variability-selected AGN from the GOODS North and
South field in the Hubble Space Telescope (HST) F850LP band. This is the first
variability-selected AGN catalog with a known expected contamination rate.
Properties of the variability-selected AGN will be discussed in an
upcoming paper.

Sample selection and data reduction are presented in Section 2. We present and
discuss statistical methods for variability selection in Section 3. We present
the application of the discussed methods to our data set in Section 4, followed by
Discussion in Section 5 and Summary and Conclusions in Section 6.

\section{Sample Selection and Data Reduction}

For our variability study we select all objects from the GOODS five epoch
$z$-band catalog \citep{giavalisco_great_2004} with a signal-to-noise of 20 or
greater. This results in a magnitude limit of about 25-26 in the $z$/F850LP band.
The $z$ band was chosen as it has the deepest imaging of all the space-based
optical observations. Another factor in choosing the most appropriate
waveband is the fact that AGN are on average more variable at shorter wavelengths
\citep[e.g.][]{di_clemente_variability_1996}. For this study, we decided to use
the $z$-band as it is considerable deeper than the redder bands. Additionally,
using blue wavebands might increase the influence of star forming regions,
especially at higher redshifts and therefore diminish our detection power. The
signal-to-noise cut-off showed to be appropriate for this study as the percentage
of variability-selected objects at magnitudes greater than 25 starts to be
extremely low. This indicates that the signal-to-noise at fainter magnitudes is
too low to pick up variability.

Data have been taken with the Advanced Camera for Surveys (ACS) Wide Field Channel
(WFC) aboard Hubble Space Telescope (HST) in the F850LP band. Data reduction is
performed using MultiDrizzle \citep{koekemoer_multidrizzle:integrated_2002}.

Photometry is performed using the NOAO-IRAF\footnote{http://iraf.noao.edu/}
aperture photometry package $daophot$. Apertures with four different radii (0.12,
0.24, 0.36 and 0.72\arcsec) centered to the catalog positions are used.
Different aperture sizes are used to determine the best aperture for
this study. Several factors are important when selecting apertures, those are:
signal-to-noise ratio, influence of small changes in PSF and influences of
galaxy light included when detecting variability of a point source against its
hosting galaxies. The smallest aperture was chosen becuase at even lower radii,
PSF changes become dominant, the largest was choosen because initial tests
showed the signal-to-noise ratio to plummet around the value choosen for the
largest aperture.

The flux in units of counts per second is derived for all five epochs
and all four apertures used. This is done by summing over the aperture and
subtracting the sky determined in an annulus with an inner radius of 3\arcsec and a width of 1\arcsec using the NOAO-IRAF sky algorithm
'ofilter'. This algorithm uses the optimal filtering algorithm and a triangular
weighting function employing the histogram of the sky pixels. It is found to
yield to show the least bias in the presence of faint objects in the
sky region (Henry C. Ferguson, private communication).

Flux measurement errors are determined using MultiDrizzle
\citep{koekemoer_multidrizzle:integrated_2002} weight maps. Inverse
variance weight maps are summed over the aperture. These measurement errors
include errors introduced by the data reduction process. To check for the correctness of the weight maps, we
perform the following test. Blank sky positions are selected and the flux in the
blank positions are measured over five epochs. The errors for the sky positions
are then determined from the weight maps. No variability is detected using this
process. We therefore conclude that the weight maps reflect the errors
from the data reduction process correctly. If the weight maps would be
incorrect, they would either over- or underestimate the flux measurement
errors. As there should be no variability in blank sky positions, this
procedure is suitable to determine if the measurement errors derived from the
weight maps are correct.

Background subtraction errors are included by quadratically adding the
standard error of the mean in the sky ring times the number of pixels in the
aperture to the flux error from the weight maps. To account for shot noise,
exposure time corrections are determined. To do this, we examine the
distributions of the flux maesurement errors from the weight maps
derived for the objects. As the error determined from the weight maps
only includes errors introduced by the data reduction process, the flux
error distribution shows three separate peaks, corresponding to areas with four,
two or one pointings per epoch (i.e. longer exposure times correspond to
lower errors from the data reduction process). According to the
measurement errors determined by the weight maps, the corresponding
exposure time is assigned for each object in each epoch. Using this exposure time, we calculate
the full standard deviation and variance from shot noise. Standard deviation and
variance are scaled to units of counts per second. The standard deviation is
calculated as follows:

\begin{equation}
\sigma_{full} = \frac{ \sqrt{ ( \sigma_{wht} \times t_{exp} ) ^ {2} + flux
\times t_{exp} }} {t_{exp}}
\end{equation}

Where $\sigma_{wht}$ is the measurement error from MultDrizzle weight map,
$flux$ is the flux in the given aperture and $t_{exp}$ is the inferred net
exposure time.

The scaled variance is calculated as follows:

\begin{equation}
var_{full} = \frac{( \sigma_{wht} \times t_{exp} ) ^ {2} + flux \times t_{exp}}
{t_{exp}}
\end{equation}

Note that we calculate both the standard deviation and the variance scaled to
units of counts per second. Due to the fact that the square root is not a linear
operator, $var = \sigma^{2}$ is only valid in units of electrons. Therefore,
when fluxes are scaled to counts per second, the corresponding standard
deviation and variance have to be calculated separately as $var = \sigma^{2}$
no longer holds.

\section{Statistical Methods for Variability Detection}

In this section, we present and discuss different statistical methods suitable
for variability detection in sparsely sampled data. These methods are compared
to methods previously used in similar studies
\citep{sarajedini_v-band_2003,cohen_clues_2006,klesman_optical_2007,trevese_variability-selected_2008}.
Statistical calculations are performed using the Python numerical and scientific
packages NumPy\footnote{http://numpy.scipy.org/} and
SciPy\footnote{http://www.scipy.org/SciPy}. NumPy random number generators are
used for all simulations\footnote{Numpy random seeds from /dev/urandom}.

For sparsely sampled data, only few statistical methods are suitable for
variability detection. In this paper, we test and discuss 
the $\chi^{2}$, $F$ and $C$ statistics.

The $\chi^{2}$ statistics is widely used in general model fitting. In
variability detection, it is used under the null-hypothesis that a flat
line fits the data, which corresponds to a null hypothesis that the object is
not variable. It is calculated as follows:

\begin{equation}
\chi^{2} = \sum_{i=1}^n \dfrac{(y_{i} - \bar{y})^{2}}{\sigma_{i}^{2}}
\end{equation}

Where $y$ is the flux, $\bar{y}$ is the mean over all $y_{i}$ and
$\sigma_{i}$ is the measurement error associated with a given flux measurement $y_{i}$. This method has the
advantage that it associates each flux measurement with its error estimate (as
we will see soon, this is not the case for the other two methods). It
is therefore well suited if one expects the errors to show significant
deviations among epochs.

The $F$ statistics compares the expected to the observed variance. It is
calculated as follows:

\begin{equation}
F = \dfrac{var_{observed}}{var_{expected}}
\end{equation}

Where $var_{observed}$ is the variance in the flux measurements and
$var_{expected}$ is the mean of the flux error estimates given as
variances. Note that, opposed to the $\chi^{2}$ statistics, the measurements
errors are not associated with individual flux measurements. Instead, both the
flux measurements and error estimates are considered as samples. The
$F$ statistics should therefore be restricted to cases in which similar errors for each measurement are expected or observed.

The $F$ statistics is also used in the ANalysis Of VAriance (ANOVA). This
statistical method is a powerful tool for variability detection in cases in which
the time domain is oversampled and many measurements are available. ANOVA does
not rely on error measurement but derives the expected variance from sub-samples
of the data. It is therefore used in micro-variability studies
\citep{de_diego_comparative_1998,villforth_intranight_2009} and is shown to have
a very high power (where power is defined as one minus the rate of Type
II errors or false negatives) \citep{de_diego_testing_2010}. ANOVA can however
not be used in this study due to the small number of data points available.

The $C$ statistics compares the expected to the observed standard deviation. It
is calculated as follows:

\begin{equation}
C = \dfrac{\sigma_{observed}}{\sigma_{expected}}
\end{equation}

In our case, $\sigma_{observed}$ is the observed standard deviation in the fluxes
and $\sigma_{expected}$ is the mean of the flux error estimates given
as standard deviations. Note that this is mathematically not identical to the
$F$ statistics where the variances are compared. From a mathematical standpoint it should be noted that opposed to the
variance, the standard deviation is not a linear operator. For example, when
adding errors, variances can be added directly while standard deviations cannot.
Just as the $F$ statistics, this method does not relate each measurement to its
error and should therefore only be used when similar errors are expected for each
epoch.

Previous studies of similar data sets have used other methods.
\cite{sarajedini_v-band_2003} analyzed two epochs of $V$-band data of the Hubble
Deep Field (HDF). They calculated the difference in magnitudes between the two
epochs and declared all objects with a deviation three times bigger than the
standard deviation of all differences at a given magnitude as variable. They
found the distribution of the differences to be close to Gaussian and therefore used
2-tailed p-values for a Gaussian distribution to estimate the number of false
positives. This method seems very robust, it also provides false positive
estimates.

\cite{cohen_clues_2006} analyzed four epochs of data from the Hubble Ultra Deep
Field (HUDF). They compared each of the measurements in the four epochs to the
measurements in all other epochs, yielding six measurement pairs per object. They
then empirically determined the error distribution for each epoch pair and
declared all objects with a difference bigger than $3\sigma$ variable for the
given epoch pair. Variability is then derived by analyzing the magnitude
differences in each of the pairs. This is comparable to the method used by
\cite{sarajedini_v-band_2003}. However, they only had two epochs of data and
therefore were limited in their choice of appropriate statistical methods.

The method used by \cite{cohen_clues_2006} has the downside that
variability is estimated from six different values and the ultimate selection
criterion used is rather complex. Determining false positive rates
would be possible using simulations, but is not straight forward due to the
complex selection criterium used.

A flavor of the $C$ statistics was used by both
\cite{trevese_variability-selected_2008} and \cite{klesman_optical_2007} . In
this method the standard deviation of the measured magnitudes of an individual
object is compared to the mean standard deviation of the entire sample under the
assumption that most objects are not variable. The two studies differ slightly in
the way they calculate the mean error for each object. While
\cite{trevese_variability-selected_2008} average over the standard deviations,
\cite{klesman_optical_2007} parameterize variability using a quantity
derived from the average of the variances in the magnitudes measured for each
epoch. While it is true that $var = \sigma^{2}$, it should be kept in mind that
a square root is not a linear operator. Therefore, $var = \sigma^{2}$ is only
valid in the space of the measurement. This equation is however no longer valid
when the measurement is transposed into another space like counts per second
(which represents a multiplicative transformation from the original space of the
measurements) or magnitude (which is a highly non-linear transformation)
\footnote{The reader can easily verify this by thinking about the variance and
standard deviation of a flux measurements with 100 counts in 10 seconds that
needs to be transferred into counts/second space. It will become clear that $var
= \sigma^{2}$ cannot be true in the transformed space.}.

In those previously mentioned studies, sources that show a scatter of
more than e.g. three times the normal scatter are then labeled variables. These methods
are similar to C statistics under the assumption that all objects at a given
magnitude have the same errors (i.e. the errors are fully shot-noise dominated
and the net exposure times are identical for all objects and epochs).

This method has several problems and caveats. Using this method, it would require
extensive simulations to associate the chosen selection criterion (in the
example, $3\sigma$) with a significance value. Such simulations have not been
performed by the authors. Therefore, no expected false positive rates are
avialable for those studies. This caveat can be important for large
samples if they have a high number of false positives, in which case it would
adversely impact the statistical significance of the number of detections. The
statement that there are 100 variables technically only means that the 100 most
variable objects are listed. The fact that the variability limit
is derived from the data also introduces the problem that more variable sources
will result in a higher limit, i.e. the detection limit depends on the number of
variable sources. This will for example result in an overdetection of variability
in faint sources, caused by the fact that this method enforces similar percentage
detection rates at every magnitude.

Additionally, using this method gives little control over the correctness of the
error measurements. While the method takes into account the real scatter of the
data, it cannot determine if the measured errors are correct. Therefore, point
spread function (PSF) changes or location dependent errors will remain unnoticed
using this method. Apprehensions concerning erroneous error estimates seem very
realistic given the complexity of the errors that can be introduced in the data
reduction process and the known complexity of space-based  PSFs.

On a side-note, using magnitudes for variability detection seems to be very
common. However, errors in magnitude space will be both
non-Gaussian and asymmetric, while being close to Gaussian in flux space. When
using common statistical estimators, this can cause problems as most statistical
methods explicitly or in-explicitly assume Gaussian errors.

\subsection{Calibrating statistical methods}

To compare the different methods, the p-values for the different estimators need
to be calibrated. p-values are values assigned to a given value of an estimator
(in our case $\chi^{2}$,$F$ and $C$). The p-value for a given value of the
estimator is the probability that a value as 'extreme' occurs from random data.
Two different p-values exist, and they should not be confused.

Two-tailed p-values give the probability that a value derived from
random data is further away from the center of the distribution than the
given value or values. Two-tailed p-values are symmetric for symmetric
distributions. For assymetric distributions they are defined such that each
tail contains the same probability.

One-tailed p-values give the probability that a value derived from
random data is bigger than the given value. This is the p-value to be used for
variability studies. It can be derived for all distributions, including
asymmetric ones.

Wheter one- or two-tailed p-values should be used depends on the
variability statistics used. For distrubutions in which both tails represent
extreme variability (for example, if deviations from catalogue magnitudes are
given), two-tailed p-values should be used. For distributions in which one tail
represents extreme variability while the other tail represents low variability,
the one-tailed p-values should be used (for example when using the $\chi^{2}$
statistics where low values indicate low variability and high values represnt
high variability).

The p-value associated with the value at which objects are considered variable is
equal to the significance. The significance determines the number of false
positives or Type I errors. It will also influence the power of
the test. The power describes the percentage of real variables detected variable,
and therefore is related to the number of false negatives (Type II errors).
Choosing a very strict significance limit will result in a low power. On the
other hand, choosing a very loose significance limit will result in a higher
power. Therefore, when comparing different methods one has to be sure to compare
them at the same significance.

Some authors use $3\sigma$ or $5\sigma$ to describe the significance
levels used in their studies, defining objects that deviate from the commonly
observed scatter by more than $3/5\sigma$ as variable. This is consistent in
itself but often wrongly related to p-values. For example, relating $3\sigma$ to
a significance level of 99.7\% is wrong in the concept of variability detection
as this is the two-tailed p-value for the Gaussian distribution.
As discussed above, using two-tailed p-values is correct only for certain
statistics. And this still leaves aside the fact that the distribution is likely
not Gaussian and therefore the association between the $\sigma$ values and the
p-values for Gaussians does not hold.

The p-values for the three statistics used in this study at different
significance levels are shown in Table \ref{p-value}.

\begin{table}
\caption{One-tailed p-values for estimators $\chi^{2}$, $F$ and $C$. P-values
are given for light curves with 5 data points at different significance
values.\label{p-value}}
\begin{tabular}{ccccc}
\tableline
Estimator & p = 95\% & p = 99\% & p = 99.9\% & p = 99.99\%\\
\tableline
$\chi^{2}$ 	& 9.48 & 13.26 & 18.46 & 23.41\\
%$\chi^{2}$ 	& 10 & 16.92 & 21.67 & 27.89 & 33.78 \\
$F$		& 1.89 & 2.65 & 3.69 & 4.70 \\
%$F$		& 10 & 1.69 & 2.17 & 2.79 & 3.36 \\
$C$     & 1.38 & 1.63 & 1.92 & 2.17 \\
%$C$		& 10 & 1.30 & 1.47 & 1.67 & 1.83 \\
\tableline
\end{tabular}
\end{table}

\subsection{Testing the power of different statistical tests}

Using the p-values derived in the previous section, the power of the different
statistical tests can be determined. Mock variable light-curves are created,
randomized and the statistical estimators are derived. Detection rates are then
calculated. Three different types of mock variable data sets are studied:

\begin{itemize}
\item \textbf{Noise}: the flux for each data points is drawn from a
normal distribution with a range of standard deviations $\sigma_{var}$
\item \textbf{Slope}: the underlying theoretical light-curve is a simple linear
trend with a range of slopes, each data point is randomized
with a standard deviation of $\sigma_{o}$
\item \textbf{Burst}: the underlying light-curve is completely
uniform with only a single outlier with a range of differences between
base and peak value, each data point is randomized
with a standard deviation of $\sigma_{o}$
\end{itemize}

A measurement error of $\sigma_{o}$ is assigned to all flux values. A range of
'variability strengths' $V$ is studied for all mock lightcurve classes. The
variability strengths $V$ are defined as follows for the different mock
light-curve classes:

\begin{itemize}
\item \textbf{Noise}: $V = \dfrac{\sigma_{var}}{\sigma_{o}}$
\item \textbf{Slope}: $V = \dfrac{max(light curve)-min(light curve)}{\sigma_{o}}$
\item \textbf{Burst}: $V = \dfrac{peakvalue - basevalue}{\sigma_{o}}$
\end{itemize}

The power of the different statistical tests (i.e. the detection rate) at a given
significance is then derived for the different types of light curves
for a range of variability strengths. The detection rates for
three methods show only minimal deviations that are consistent with error
expected from the fact that we only use two decimals and therefore our
p-values are not completely accurate. The results are shown in Table
\ref{power}.

\begin{table}
\caption{Detection power (in per cent) for different mock light curves at a
significance level of 99.9 per cent. $V$ is the strength of the
varaibility as defined in the text.
\label{power}}
\begin{tabular}{c|ccc}
\tableline
$V$ & Slope & Burst & Noise \\
\tableline
0.5 & 0.11 & 0.17 & 0.00 \\
1.0 & 0.20 & 0.35 & 0.14 \\
1.5 & 0.40 & 0.97 & 8.33 \\
2.0 & 0.82 & 2.46 & 32.96 \\
2.5 & 1.59 & 5.74 & 56.55 \\
3.0 & 2.94 & 12.14 & 72.53 \\
3.5 & 5.57 & 22.62 & 82.59 \\
4.0 & 9.65 & 37.15 & 88.45 \\
4.5 & 15.57 & 53.80 & 92.12 \\
5.0 & 23.64 & 69.91 & 94.70 \\
5.5 & 33.32 & 82.97 & 96.32 \\
6.0 & 45.39 & 91.74 & 97.23 \\
6.5 & 56.99 & 96.60 & 97.93 \\
7.0 & 68.56 & 98.81 & 98.40 \\
7.5 & 78.35 & 99.62 & 98.83 \\
8.0 & 85.97 & 99.92 & 99.05 \\
8.5 & 91.90 & 99.98 & 99.28 \\
9.0 & 95.63 & 99.99 & 99.38 \\
9.5 & 97.88 & 100.00 & 99.54 \\
\tableline
\end{tabular}
\end{table}

\subsection{Testing robustness of tests}

As we have shown, the three different statistical tests have equal power
for 'perfect' data. However, real data is hardly perfect. Therefore, we will
determine how robust the statistical tests are in presence of deviations from the perfect simulated data.

In the previously presented test, the flux measurement errors (meaning
the standard deviation or variance) are measured accurately. However, this is not expected for real data. Errors are
measured in a similar way to fluxes and this process introduces measurement
errors also into the error measurement. Therefore, it is of interest to see how
different tests perform for 'erroneous' error measurements. For our mock data
that means that the flux measurement errors will also be drawn from a
normal distribution with a certain width. The width of this distribution will give the 'defectiveness'
of the error measurements.

When rerunning the tests using 'erroneous' errors we find that the different
tests differ in their detection power. The $\chi^{2}$ statistics now shows the highest
detection power, the $C$ statistics shows the second highest power and the
$F$ statistics shows the lowest power. Results for the detection power with
'erroneous' errors are shown in Table \ref{power_err}, located in the Appendix.

We  therefore conclude that the $\chi^{2}$ statistics should be used in cases in
which the errors show strong deviations between the different epochs. The $C$
statistics is to be preferred over the $F$ statistics due to its greater power
under the influence of erroneous error measurements. We will therefore only use
the $C$ and $\chi^{2}$ statistics for our study.

\subsection{Estimating the influences of sparse sampling}

Additionally, it is of interest to understand the detection power for AGN
light curves. To determine the detection power, we create mock AGN light-curves
using the method introduced by \cite{timmer_generating_1995}. This
method randomizes both the phase and the amplitude of the Fourier transform. In other
methods, only the phase is randomized and therefore only a subset of all
possible lightcurves is simulated \citep{timmer_generating_1995}.

\begin{deluxetable}{cccc}
\tabletypesize{\scriptsize}
\tablecaption{Detection power (in per cent) for mock AGN light curves for
different significance levels and different errors in per cent as
defined in the text.
\label{power_AGN}}
\tablehead{
\colhead{error in \%} &
\colhead{Power (95\%)} &
\colhead{Power (99\%)} &
\colhead{Power (99.9\%)}
}
\startdata
0.11 & 100.00 & 100.00 & 100.00 \\
0.23 & 100.00 & 99.97 & 99.97 \\
0.34 & 99.97 & 99.97 & 99.95 \\
0.45 & 99.95 & 99.95 & 99.93 \\
0.57 & 99.93 & 99.92 & 99.83 \\
0.68 & 99.92 & 99.82 & 99.52 \\
0.79 & 99.82 & 99.53 & 99.29 \\
0.91 & 99.56 & 99.32 & 98.87 \\
1.02 & 99.44 & 99.02 & 98.25 \\
1.13 & 99.22 & 98.73 & 97.40 \\
2.27 & 91.99 & 87.42 & 81.46 \\
3.40 & 78.12 & 69.34 & 60.51 \\
4.54 & 63.25 & 53.08 & 42.69 \\
5.67 & 49.47 & 38.72 & 28.92 \\
6.81 & 37.74 & 27.84 & 18.82 \\
7.94 & 28.81 & 19.34 & 11.81 \\
9.08 & 21.23 & 13.33 & 7.32 \\
10.21 & 15.58 & 8.86 & 4.11 \\
11.35 & 10.92 & 5.49 & 2.13 \\
12.48 & 7.80 & 3.32 & 1.21 \\
13.62 & 5.24 & 1.93 & 0.53 \\
14.75 & 3.36 & 1.18 & 0.28 \\
15.89 & 2.09 & 0.61 & 0.13 \\
17.02 & 1.43 & 0.35 & 0.03 \\
18.16 & 0.88 & 0.20 & 0.00 \\
19.29 & 0.43 & 0.07 & 0.00 \\
20.43 & 0.33 & 0.02 & 0.00 \\
21.56 & 0.19 & 0.00 & 0.00 \\
\enddata
\end{deluxetable}

10000 mock AGN light curves are created. We simulate light curves ten times
longer than the sampling time-scale to include the red noise leak \citep[see
e.g.][for a discussion of the red noise
leak]{vaughan_characterizingvariability_2003}, five data points are drawn from
the mock light curves with sampling similar to the GOODS five epoch time
sampling. The data are then randomized with a range of errors and data are
analyzed using the three statistical estimator discussed.

As a measure of the variability strength, we give the ratio between the
assigned flux measurement errors and the median spread in the analyzed
light curves in percent. With changes on a timescale of years of typically around 1 mag in AGN light curves, a
common error of 0.1 mag would result in a signalstrength of about 3, resulting
in a detection probability of about 78\% at 95\% significance. Detection
probabilities for other typical errors and variability strengths can be derived from Table
\ref{power_AGN}.

\section{Results}

\subsection{Zero-point Calibration}

First, we check for possible zero-point offset between epochs. We
analyze offset in the zero-point for bright, point-like objects.
Zero-point offsets are derived for aperture corrections (i.e. the
correction applied to determine the entire flux of the object instead of the flux
in the aperture) and inter-epoch zero-point drifts. This is done by calculating
the mean offset between the measured magnitude and catalogue magnitude for
bright, point-like objects. The zero-point drifts for the large apertures are
very mild, indicating changes below 0.01\%. This implies that the photometric
calibration is extremely accurate, the differences are within the error
of the estimator used.

For the smallest aperture (0.12\arcsec radius) however, the fitting indicates
inter-epoch changes on the level of $\sim$ 1\%. We therefore correct the fluxes
and errors with the derived zero-point offsets for the 0.12\arcsec aperture and
compare the number of detections with and without the correction applied. The
number of variables increases when correcting for zero-point drifts indicating
that either the error of the fitting routine is dominating or that changes in the
PSF are location-dependent. Therefore, we test for possible problems with PSF
instabilities in the next subsection.

While we could simple use the biggest aperture to avoid such problems,
this is not the optimal solutions. Big apertures can result in both low
signal-to-noise ratios and lower detection rates for faint AGN due to high
contribution from the hosting galaxy. Therefore, we will try to assess which
aperture is optimal for this study.

\subsection{PSF Stability}

Using TinyTim \citep{krist_simulation_1995}, we test possible influences of
defocusing over the field of view and changes in the PSF shape to get an
estimate of the errors expected from PSF changes.

First, images are checked for possible changes in the apparent position
between the different epochs. Objects show decentering between different epochs only on sub-pixel scales, as
this is comparable to the accuracy of the centering algorithm, no decentering is
assumed between epochs.

To test for possible errors due to changes in the PSF, PSFs for 64 ACS WFC
positions with filter F850LP are created. Defocusing is introduced with values
between -5 and +5 $\mu m$ in steps of 1. These values describe the
movement of the secondary mirror with respect to the primary, where 0 corresponds
to the telescope being in focus. Those values are typical focus changes due to
'breathing' of the HST spacecraft \citep{di_nino_hst_2008}. Both a E galaxy
template and a QSO template are used for the objects spectral shape, but no
difference show in the resulting PSFs. Aperture photometry is then performed on
each image with the same aperture sizes used for the data. From this measurement,
typical errors due to both changes of the position of the object on the chip and
the focus can be estimated. For the smallest aperture (0.12\arcsec),
typical values for PSF changes expected are $\sim 1\%$, dropping to
$\sim 0.25\%$ for the 0.24\arcsec aperture, $\sim 0.15\%$ for the 0.36\arcsec
aperture and $\sim 0.1\%$ for the 0.72\arcsec aperture.

Note that there is a known red halo in the ACS WFC for the F850LP
filter which is not included in TinyTim \citep{sirianni_photometric_2005}. This might slightly alter
our results, making the possible changes due to defocusing and changes over the
field of view smaller. False variability due to the changes in the red halo are
only expected if spectral changes occur. In case the spectral shape changes, the
object is intrinsically variable. Therefore, the red halo might cause
true variability to be boosted, but no excess false variability is expected to
be introduced due to this effect.

\begin{figure}
\includegraphics[width=8cm]{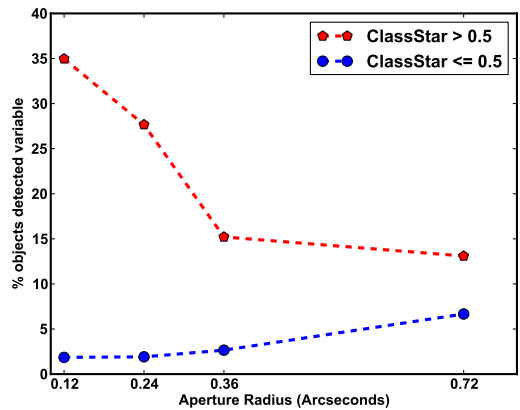}
\caption{Percentage of objects detected variable, separated into point-like
(red pentagons) and extended (blue circles) objects. Detections are performed
using $C$-Statistics for four different aperture radii.}
\label{ap_detects}
\end{figure}

Given the expected errors due to PSF changes, we will first asses if PSF
instabilities induce false variability for small apertures in the data. Figure
\ref{ap_detects} shows the detection rates for point-like and extended objects over the
aperture radius used. As we can see the detection rate for point-like objects is generally
significantly larger than for extended objects. This indicates that point-like
objects are intrinsically more variable than extended objects. This is expected
given the fact that most high-luminosity AGN would appear
point-like due to the fact that the AGN significantly outshines the galaxy.

\begin{figure}
\includegraphics[width=8cm]{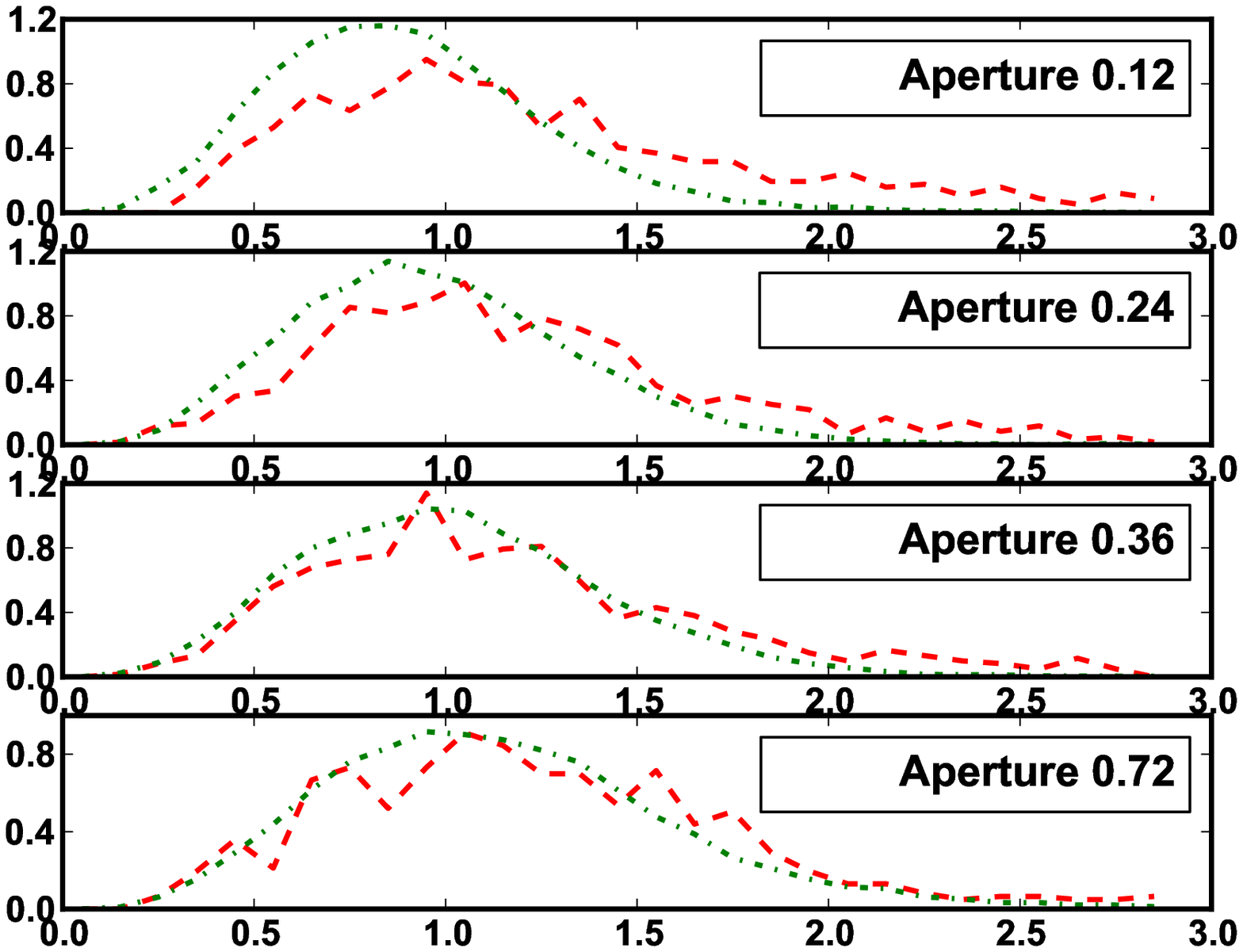}
\caption{Observed distribution for variability estimator $C$ for point-like
(red dashed line) and extended (green dash-dotted line) objects. Objects are
divided into classes according to their stellarity measure ClassStar at a value
of 0.5.}
\label{ap_comp}
\end{figure}

However, the detection rate for stars is significantly higher at the two
smallest apertures, indicating that PSF changes induce false variability. Figure
\ref{ap_comp} shows a comparison between the observed estimator
distributions of point-like and extended sources. The shape of the distribution differs
strongly for the two smallest apertures. For the two biggest apertures however,
the distributions agree well. This indicates that PSF changes will not affect
those apertures.

On the other hand, for the purpose of studying variable AGN, it should be noted
that making the aperture bigger will start to drown the variability in the
host galaxy light. This is due to the fact that bigger apertures will include
more emission from the hosting galaxy. Therefore, the smallest possible aperture
should be used. As the two smallest apertures show signs of variability due to PSF errors, we use an
aperture with a radius of 0.36 \arcsec.

\subsection{Re-calibration of statistical estimators}

The photometric errors of ACS have been studied excessively
\citep{sirianni_photometric_2005}. However, given the extreme importance of
exact error measurements in this study and the general complexity of photometric
errors, we will check and if necessary correct the error measurements derived.
To check the quality of our measurement errors ($\sigma_{full}$ as defined in the
text), we compare the distributions of measured $\chi^{2}$ and $C$ to the
theoretical distributions. It is assumed that most objects are not variable and
therefore the observed distribution should follow the theoretical distribution.

\begin{figure}
\includegraphics[width=8cm]{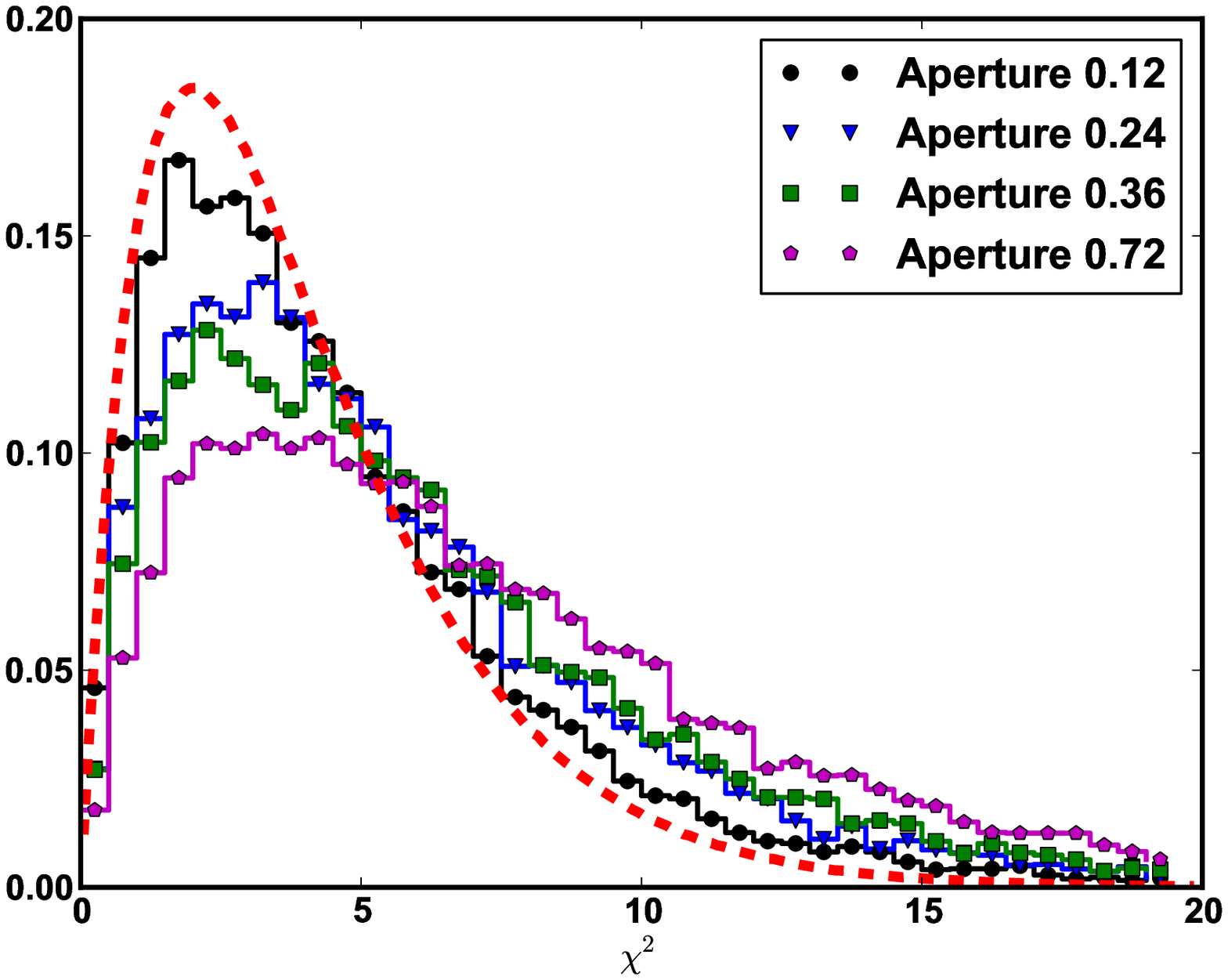}
\vspace{0.5cm}
\includegraphics[width=8cm]{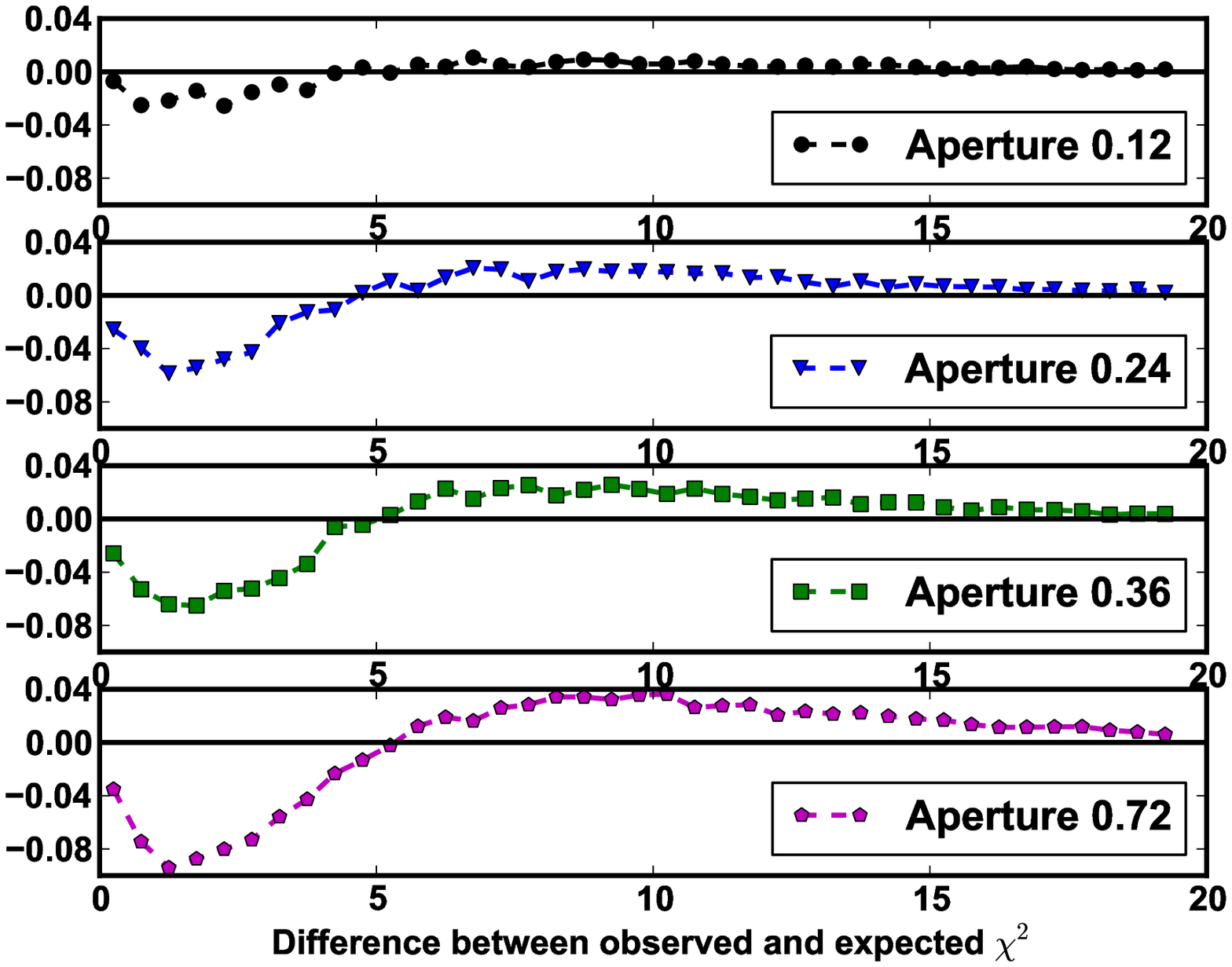}
\caption{Comparison between observed and theoretical
$\chi^{2}$ distribution for four different aperture radii. Upper panel:
theoretical distribution (red dashed line) and observed distributions (solid
line histograms). Lower panel: difference between theoretical and observed
distributions. Only objects with detections in 5 epochs are included.}
\label{rawdists_chi}
\end{figure}

We find that the observed distributions for our measurements with an aperture
radius of 0.12 \arcsec agree rather well with the theoretical distributions (Fig.
\ref{rawdists_chi}), even given the PSF problems. Using bigger apertures, the
distributions start to deviate from the theoretical ones, showing a flatter tail
towards higher values than expected (Fig. \ref{rawdists_chi}). This indicates
that either the shot noise or the background contribution are underestimated.
This will result in excess false positives. However, as we have
discussed in the last paragraph, aperture radii of 0.36\arcsec should be used to avoid false positives due to PSF changes.

\begin{figure}
\includegraphics[width=8cm]{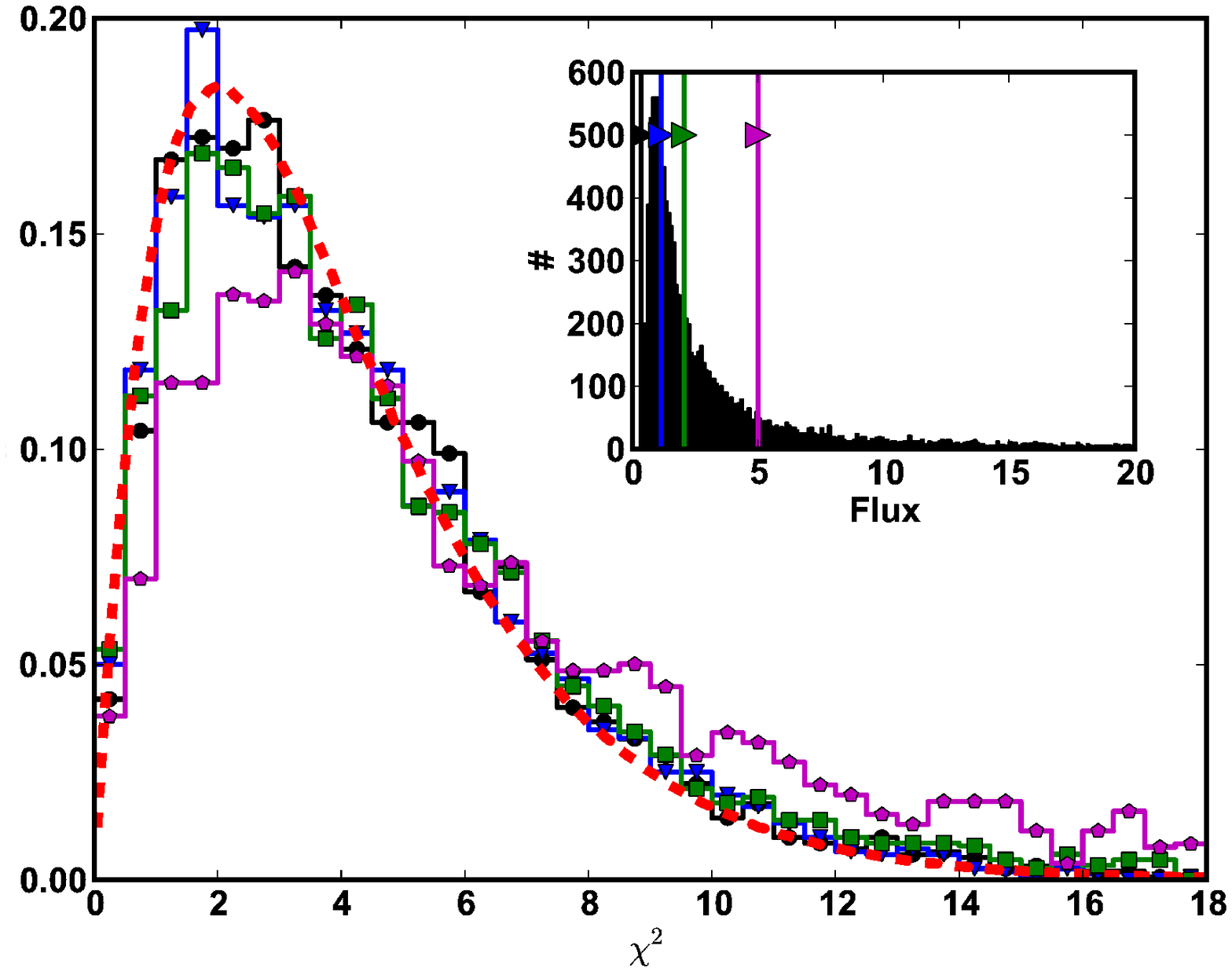}
\vspace{0.5cm}
\includegraphics[width=8cm]{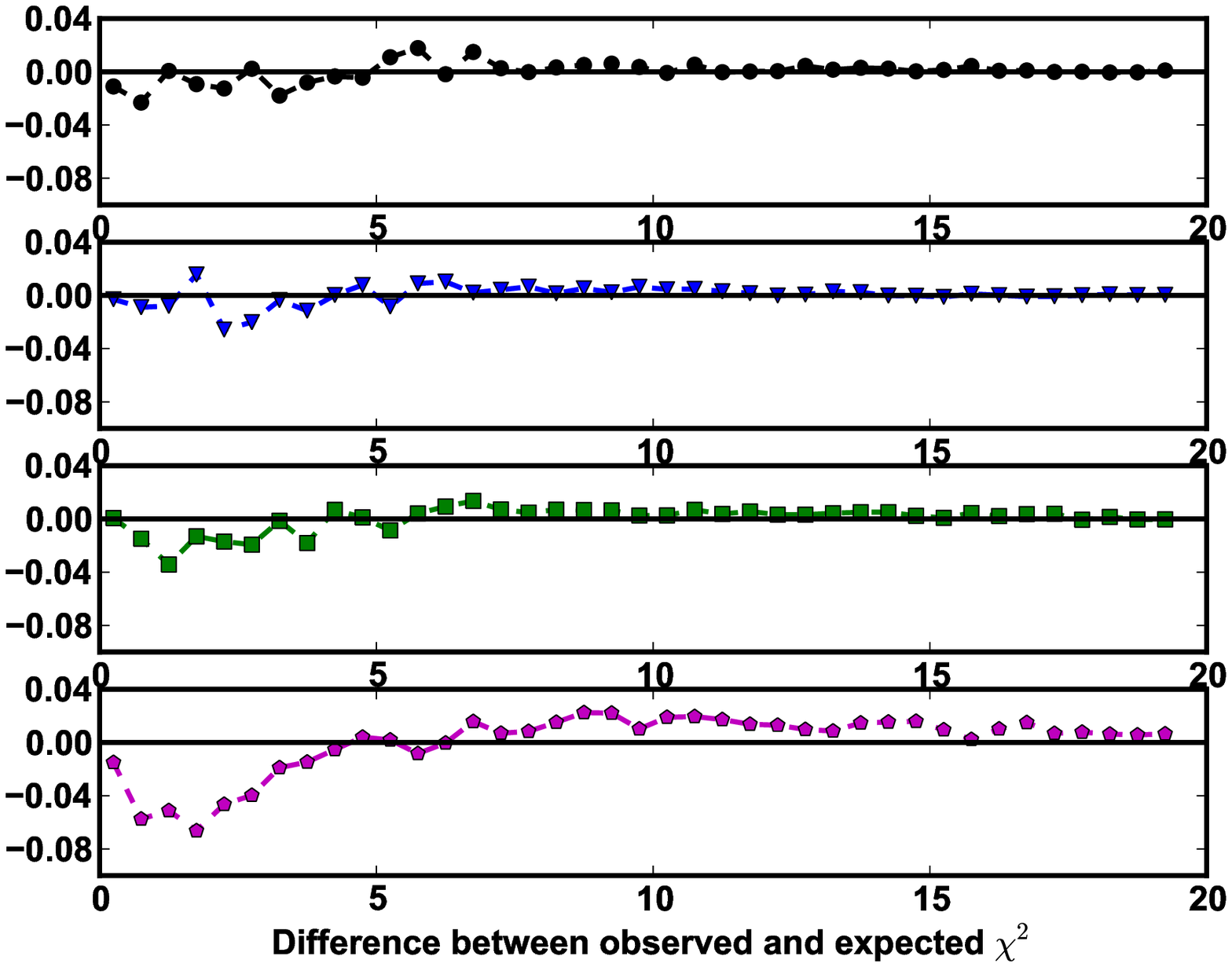}
\caption{Comparison between theoretical and observed $\chi^{2}$ distributions
for a 0.12\arcsec aperture in different flux bins. Upper panel:
theoretical distribution (red dashed line) and observed distribution (solid line histograms). The inset plot show the histogram
of catalog fluxes, the vertical lines denote the left limit flux of the
histogram of the given color. Lower panel: difference between
theoretical and observed distributions for same flux bins as in the upper
panel.}
\label{rawdists_chi_flux}
\end{figure}

\begin{figure}
\includegraphics[width=8cm]{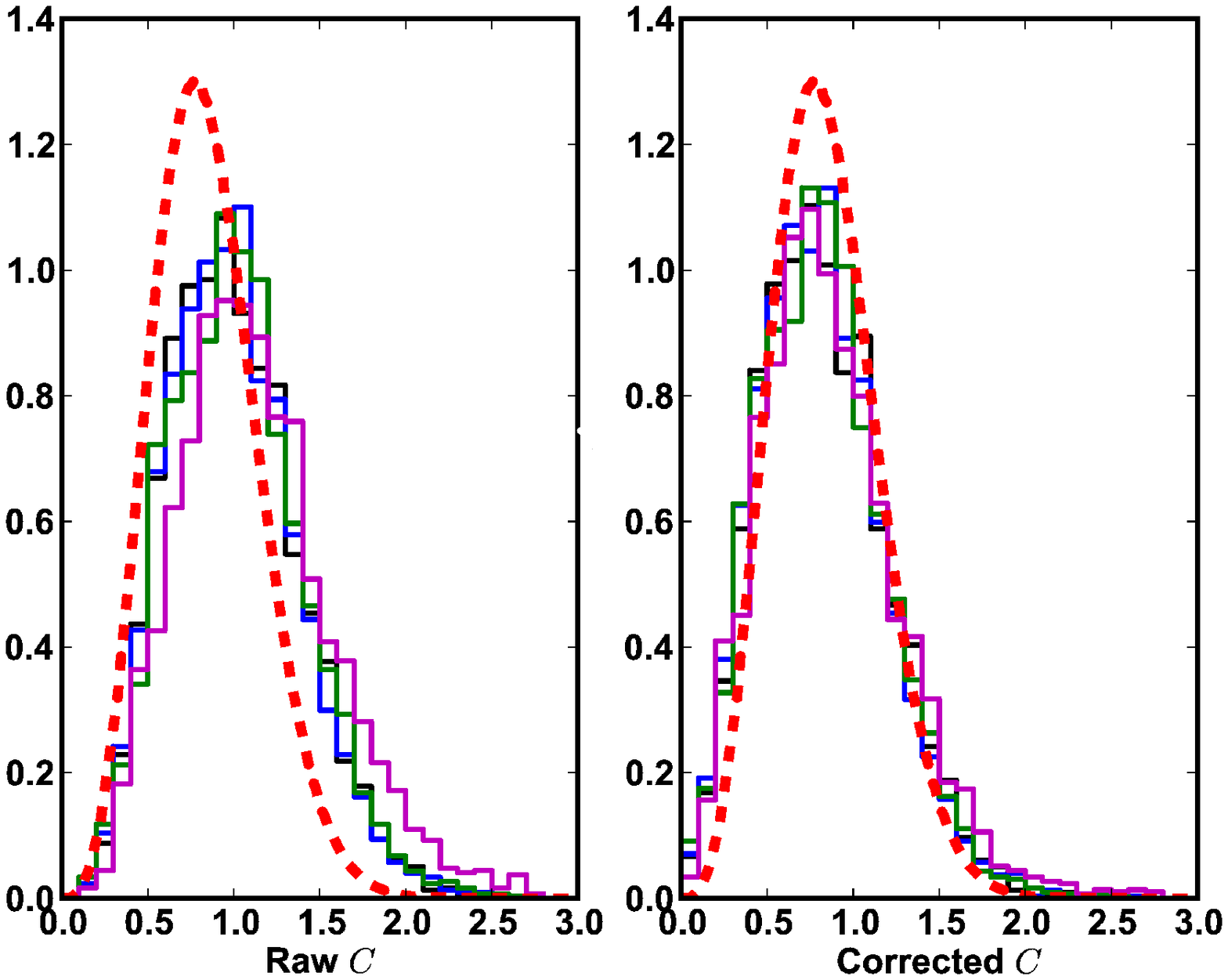}
\vspace{0.5cm}
\includegraphics[width=8cm]{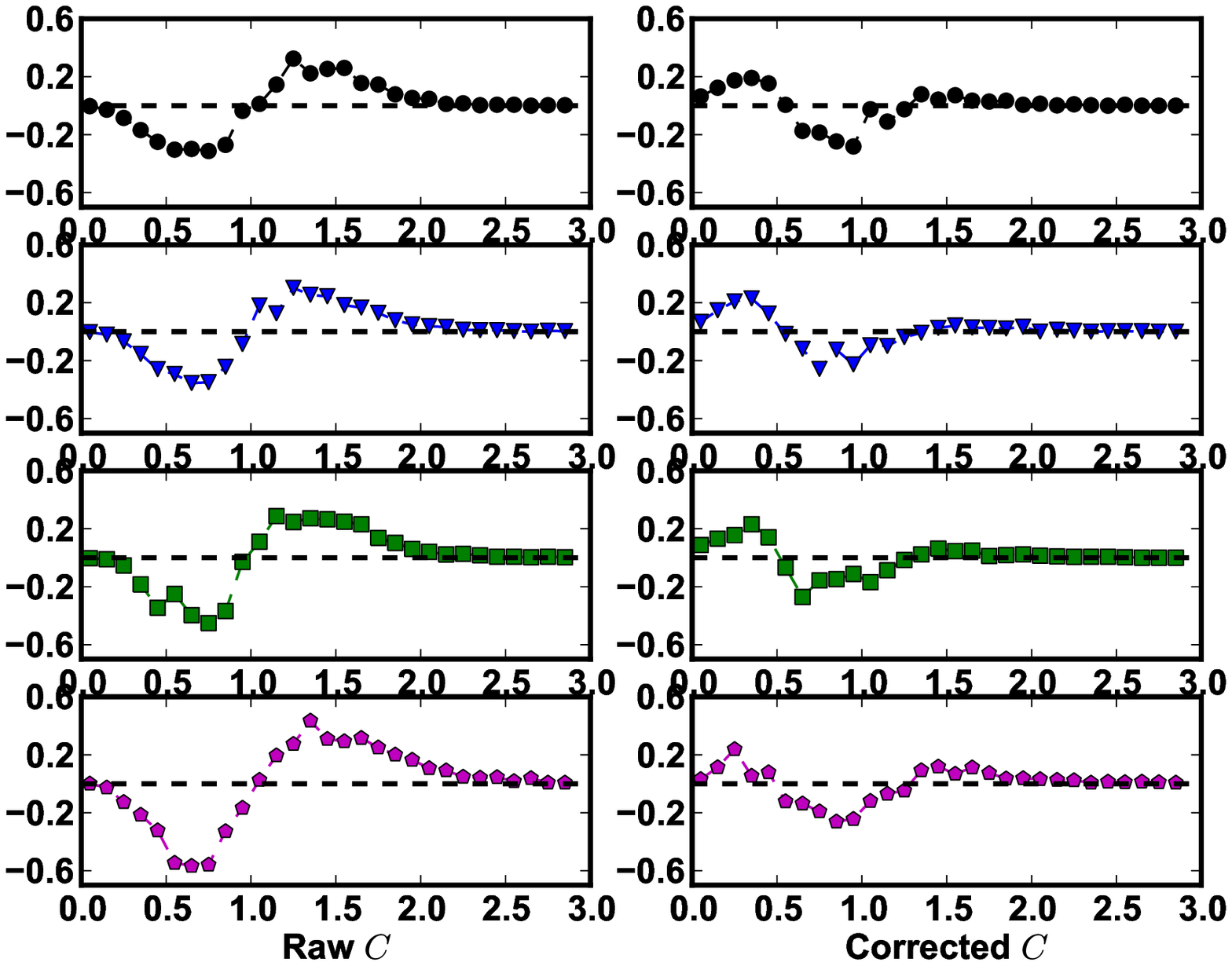}
\caption{The $C$ distribution (0.36\arcsec aperture) for four different flux
bins before (left) and after (right) the flux-correction is applied. The solid-line histograms show the
histograms for the different flux bins, the dashed red line shows the theoretical
distribution. Lower panels shows the differences between the theoretical and
observed distributions. For the meaning of the histogram colors, see legend and
inset plot in Fig. \ref{rawdists_chi_flux}}
\label{C_corrected}
\end{figure}

To check more closely for possible problems in the error determination, we
compare the variability estimator distributions to the theoretically expected
distributions for different object magnitudes. We find that the distributions
shift to higher values at higher fluxes for all apertures (see Fig.
\ref{rawdists_chi_flux} for an aperture radius of 0.12\arcsec and left panel of
Fig. \ref{C_corrected} for an aperture radius of 0.36\arcsec). This could be
related to an underestimation of shot noise errors or other factors. To correct
for this problem, which could cause an over-detection of variability in bright
sources, we decide to include a flux-dependent factor in the variability
estimators. To derive this factor, we use the $C$ statistics. When applying a
correction to the estimator, the error estimates are changed for all five epochs.
This means that a correction is applied to the sample of errors. This is more
compatible with the $C$ than the $\chi^{2}$ statistics. Therefore, from now on
the $C$ statistics will be used.

\begin{figure}
\includegraphics[width=8cm]{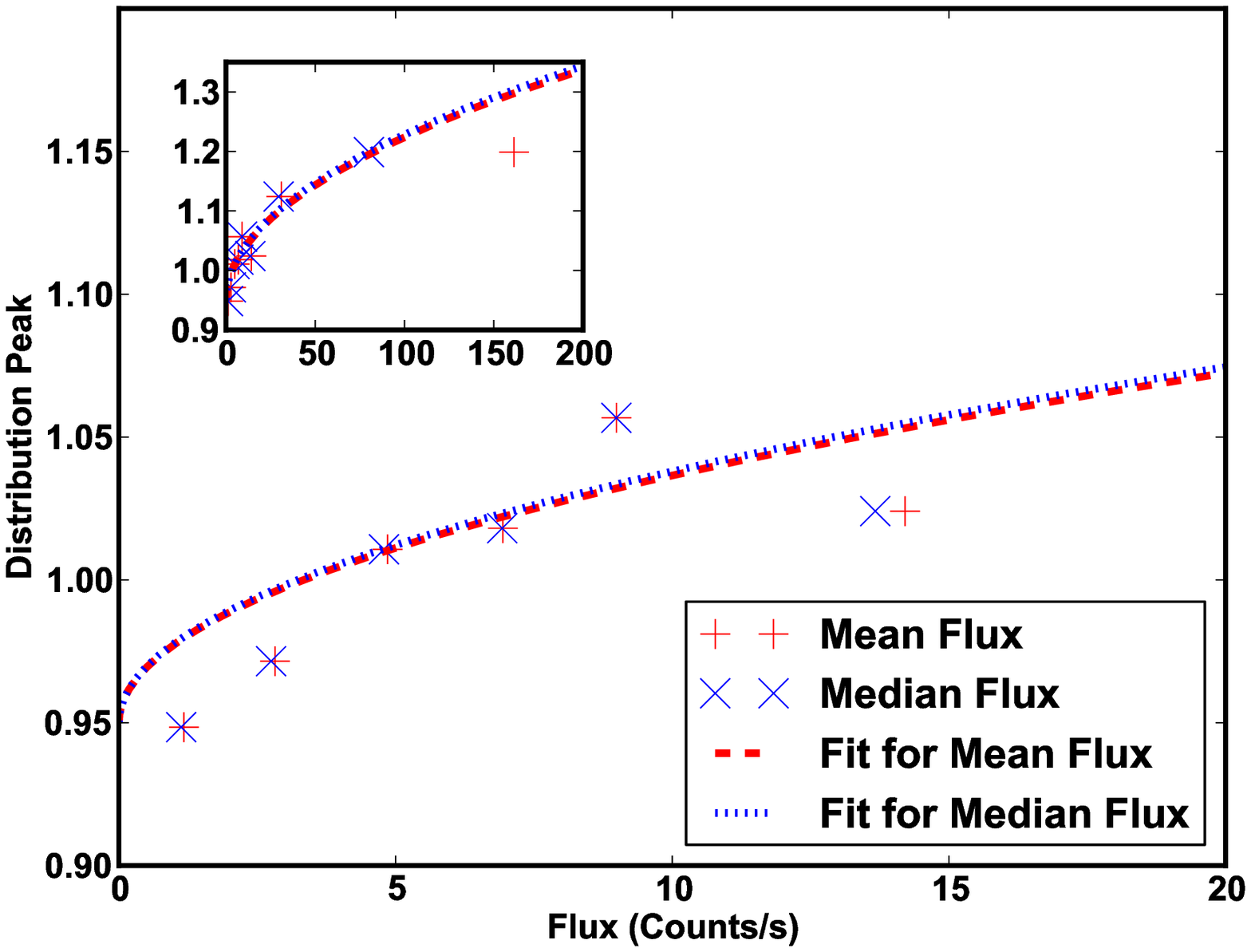}
\caption{Determining the flux-dependent error corrections for the observed
$C$-distribution. Shown is the observed peak of the $C$-distribution versus the 
mean (red + symbols) and median (blue x symbols) flux in each bin. The best fit
to the data is shown as a dashed red line for the mean and a blue dotted line for
the median. The big plot plot shows the data for low fluxes, the small inset shows
the same data for the entire flux range. The axes labels for the
inset plot are identical to the big plot.}
\label{flux_correction}
\end{figure}

To derive the correction factor, we divide the sample into seven flux bins.
A histogram of the test statistics is then calculated for each bin and
a gaussian is fitted to determine the peak position of the distribution. This estimate for the distribution peak is
then plotted against both the mean and the median of fluxes in each bin (see Fig.
\ref{flux_correction}). A clear trend for the peak to move to higher values at
higher fluxes is visible. As the increase is presumably due to an underestimated
flux error, a square root function with an y-axes offset is fitted to the data.
The correction is then applied to the data by normalizing the test
statistics with the value derived. From the definition of the $C$-Statistics,
we can see that this can also be interpreted as a correction applied to the
error estimates.

The flux corrected $C$-distributions are shown in the right panel of Fig.
\ref{C_corrected}. As we can see the distributions now agree well with
the theoretical distributions. There are small deviations, but those indicate
that the correction produced a slight overshoot, resulting in an overcorrection
for very low $C$ values. This could possibly cause an increase in the number of
false negatives. However, as the deviations are only apparent in the left tail
of the distribution, the possible effects of this overcorrection should be
minimal. Using the flux-corrected 0.36\arcsec $C$ value gives us 173 variable
sources out of 11931 total sources at a significance level of 99.9\%.

\subsection{Clustering of variable sources: testing for locally
underestimated errors}

\begin{figure}
\includegraphics[width=8cm]{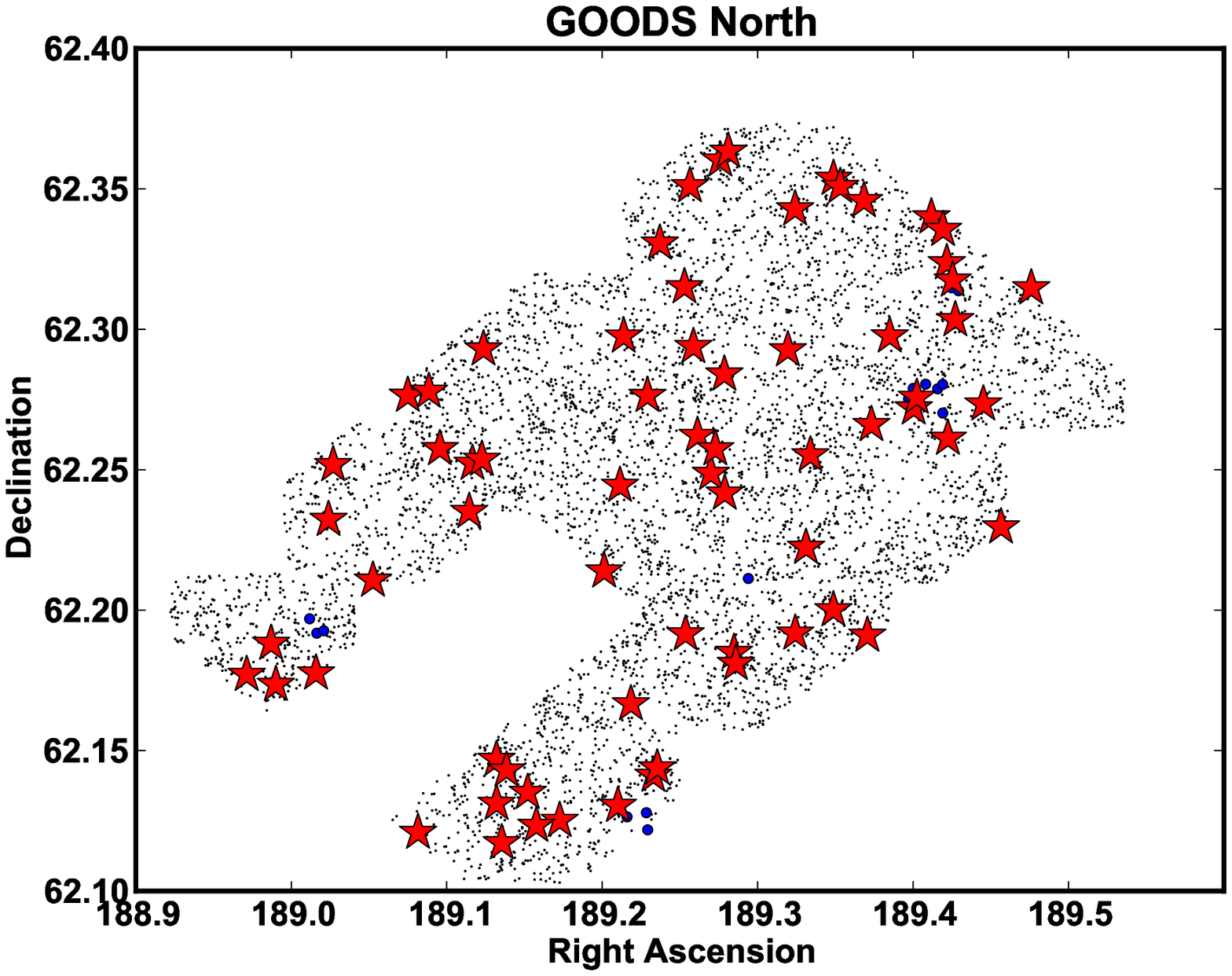}
\caption{Location of the variable (big red stars) and non-variable (small
black dots) objects in the GOODS North field. Objects that have been
rejected due to influences of nearby bright stars are plotted as small blue
circles. Only objects with detections
in all 5 epochs are included.}
\label{north}
\end{figure}

\begin{figure}
\includegraphics[width=8cm]{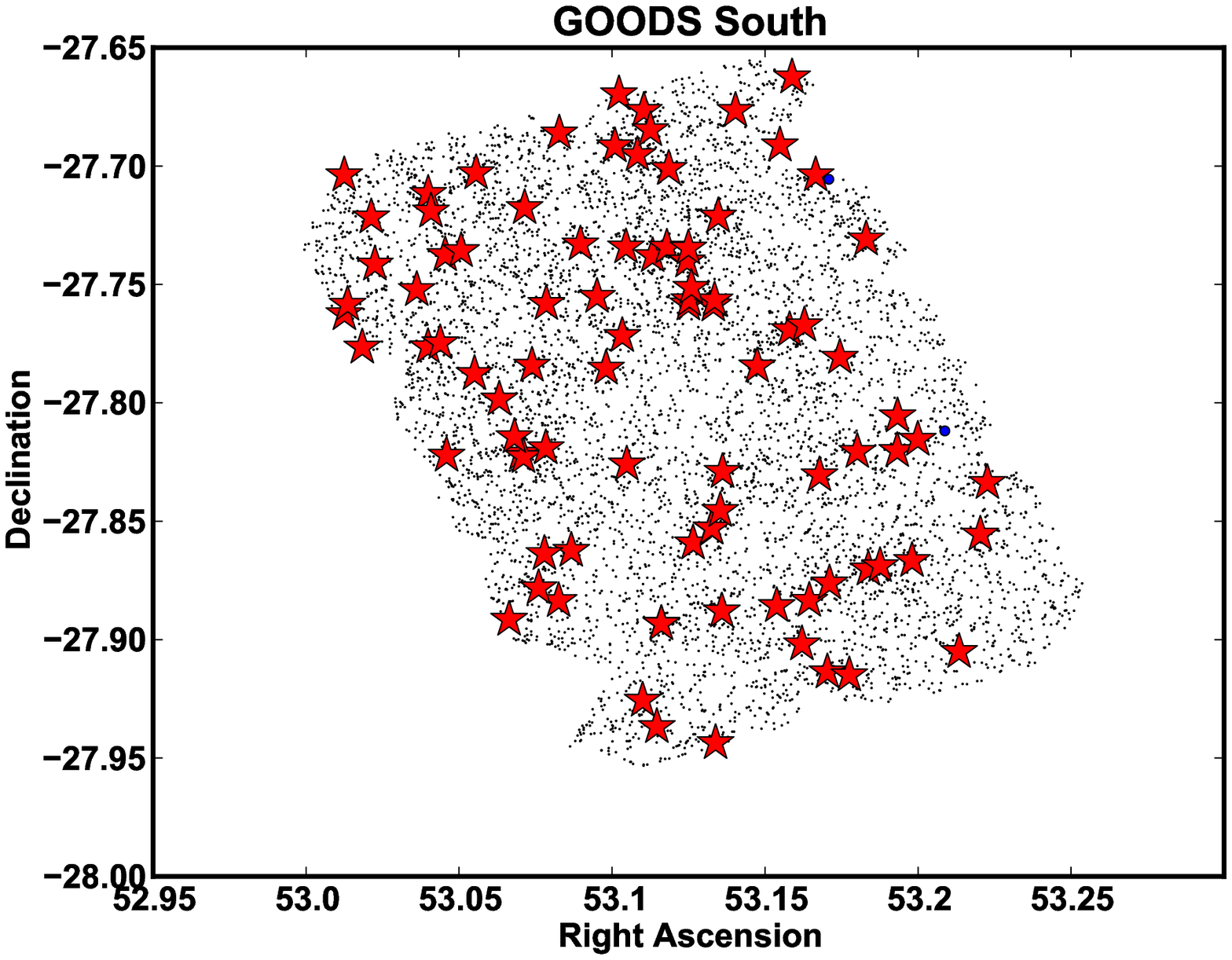}
\caption{Location of the variable (big red stars) and non-variable
(small black dots) objects in the GOODS South field. Objects that have been
rejected due to influences of nearby bright stars are plotted as small blue
circles. Only objects with detections in all 5 epochs are included.}
\label{south}
\end{figure}

Next, we will determine if location dependent errors or strongly location
dependent changes in the PSF have been underestimated. Extreme clustering of
variable sources might indicate such problems. These kind of problems could be
very hard to pick up using the previously described test as they would
potentially affect only a small percentage of sources.

The spatial distribution of the variable objects in the GOODS North and South
field is shown in Figures \ref{north} and \ref{south}. At first visual
inspection, a clear clustering at a few distinct locations is visible, especially
in the GOODS North field. We inspect these areas by eye and find that those
objects are located very close to bright stars that show clear diffraction spikes. The
light curves show heavy outliers in some epochs. Indeed, visual inspection shows
that diffraction spikes are present close to the centers of these objects. We
reject the 18 objects affected by this problem, dropping our number of variables
from 173 to 155.

However, there might still be more subtle clustering caused by underestimated
location-dependent errors that is not picked up by eyeball inspection. Therefore,
we calculate the distance to the nearest variable neighbor for all variables in
the North and South field. We then create a mock variable catalog by randomly
selecting the same number of sources for each field. Only objects with detections
in five epochs are used for the selection. The distance to the nearest neighbor
is also calculated for the mock catalogs.

The distributions of nearest neighbor distances for the variable and mock
variable catalogs are then compared using a two-sample Kolmogorov-Smirnov test.
In case the variable sources would be extensively clustered, the p-values should
be small, indicating that the samples are not drawn from the same parent
population.

We create 100000 mock catalogs for each of the fields. Kolmogorov-Smirnov tests
are then performed for the variable sample against each of the mock samples and
against a master distribution created by merging all mock distributions. The
p-values for the North field range from 0.02 to $>$ 99.9 per cent with a mean of
36 per cent and a median of 30 per cent. Comparing the distribution for the
variables to the master distribution gives a p-value of 28 per cent. The p-values
for the South field range from $10^{-4}$ to $>$ 99.9 per cent with a mean of 32
per cent and a median of 28 per cent. Comparison with the master distribution
gives a p-value of 20 per cent. This shows that no excess clustering is found
after rejecting objects affected by diffraction spikes. This finding indicates
that location dependent errors are well accounted for.

Auto- or 2-point-correlation functions might be more appropriate to study
possible clustering, but as our test with nearest neighbor distances does not
show any abnormalities, this easy test should be enough to check for problems
with location-dependent errors or PSF changes. Additionally, weak 'real'
clustering in our variability sample might be present. We will therefore not
further explore this topic.

\subsection{Rejection of saturated objects, stars and supernovae}

\begin{figure}
\includegraphics[width=8cm]{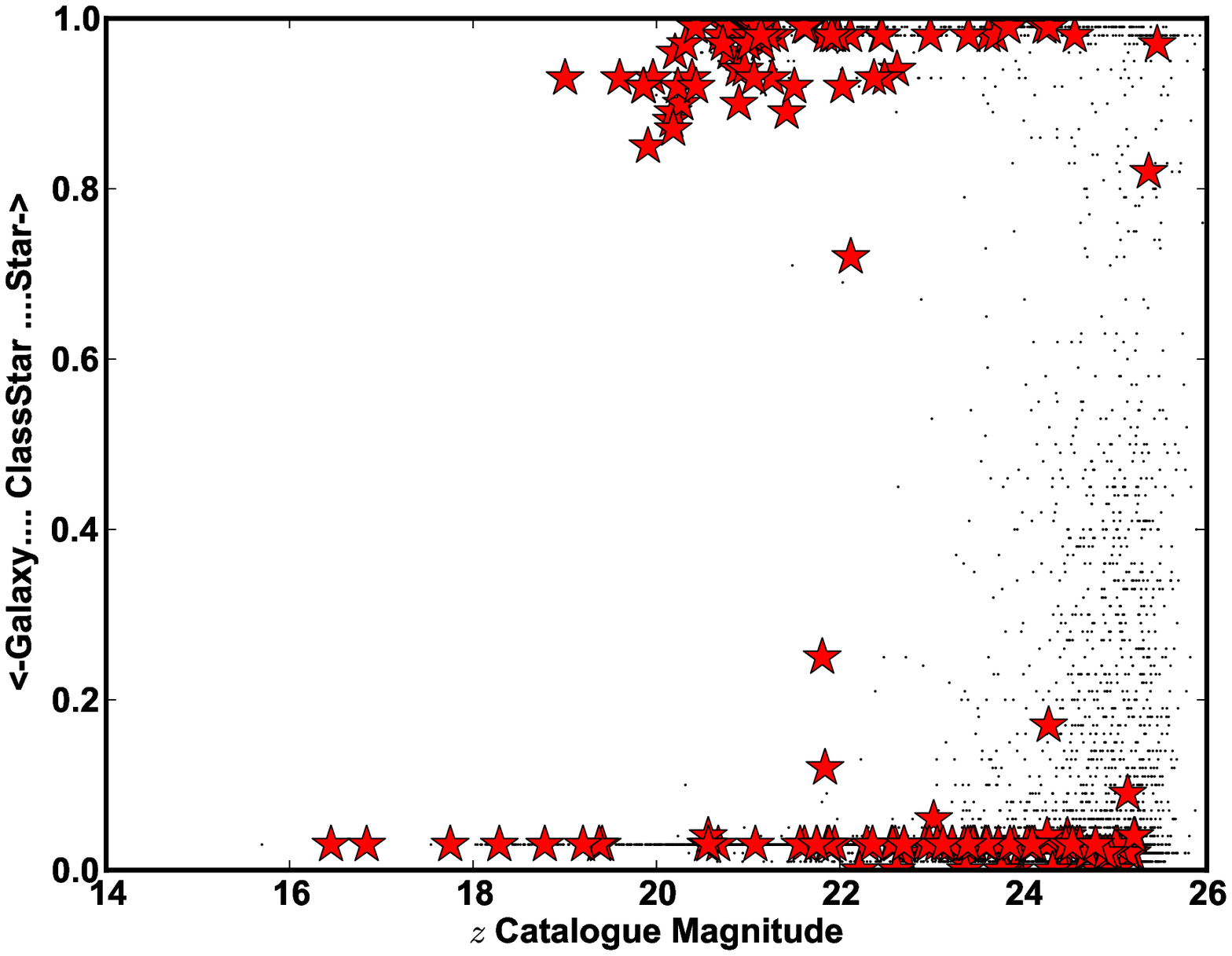}
\caption{Sextrator stellarity parameter ClassStar plotted against $z$ band
catalog magnitude. Non-variable objects are shown as small black dots,
variable objects are shown as big red stars. Objects that
have been rejected due to the influence of nearby bright stars are not shown.}
\label{cs}
\end{figure}

At this point, our catalog is only a catalog of variable objects, we are
however interested in variable AGN. To produce a catalog of variable AGN, we
need to reject saturated objects, stars and supernovae. 

Fig. \ref{cs} shows the location of the variable sources in a
magnitude-stellarity plot. As a measure of stellarity, we use the
Sextractor parameter ClassStar provided in the GOODS catalogue
\citep{dickinson_great_2003}. Objects with high stellarity can have two
potential problems. They could be saturated or the non-linear range of the chip
could be reached, in both cases, false variability is expected.

To flag possible saturated objects or those in the non-linear regime, we
determine at which magnitude the objects start to enter the non-linear regime.
Therefore, the object flux is compared to the objects peak flux for all
point-like sources and the turn-over point is marked. This happens at 18th magnitude. All
objects showing a peak flux higher than the turnover point or are
brighter than 18th magnitude and point-like are therefore rejected from our
sample. None of our detected variable objects is in this regime.
All bright stars were rejected before due to extremely high error bars, caused by
the saturation or diffraction spikes.

As for excluding stars, we have to take into account that both stars and AGN can
appear point-like. High-luminosity AGN can be more than 100 times brighter than
their hosting galaxy. Therefore, they appear point-like even if the host
galaxy could theoretically be resolved. Therefore, we will not reject point-like
objects per se.

However, we correlate our data set with other catalogs to exclude stars. 15
objects in our variable catalog turned out to be red stars and are therefore
rejected from the variable AGN catalog. They are given in the final table but
are flagged as stars. More of our variability selected objects might be
stars, but we are confident that we were able to flag the majority of stars.

Supernovae are the only other objects that show variability on the sampled
timescale. Therefore, we correlated our data with supernovae identified by
\cite{riess_type_2004}. This study identified supernovae from the five epoch
GOODS data. We found that one supernova from \cite{riess_type_2004} is identified
in this study as variable. This supernova (2003XX) went off in the very center of
an elliptical galaxy. The light curve of this object indeed shows a singly
outlier. This object is therefore rejected from the final variable AGN
catalog. It is however listed in the variable table and flagged.

\subsection{Variable AGN catalog}

\begin{figure}
\includegraphics[width=8cm]{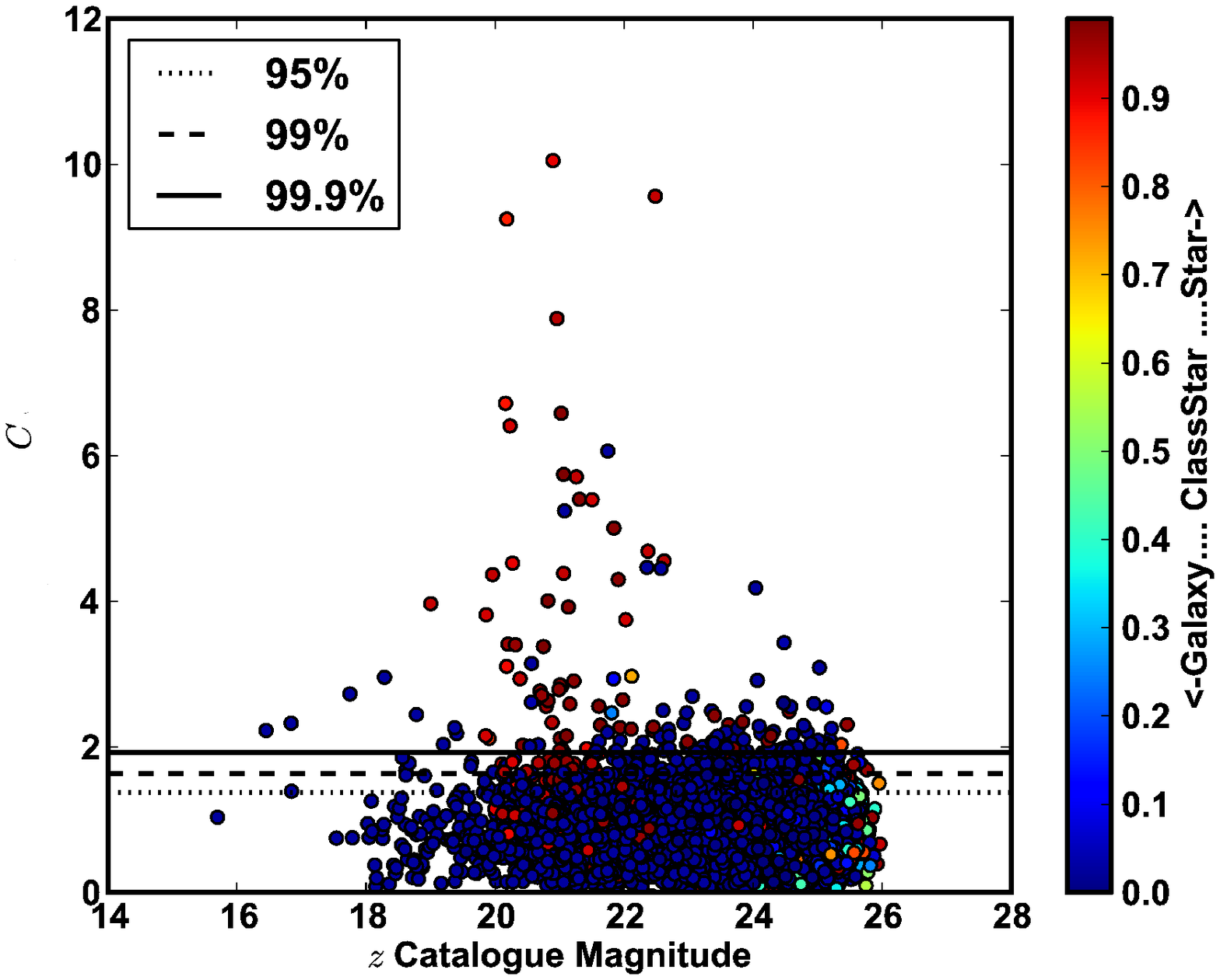}
\caption{Flux corrected variability estimator $C$ plotted against object $z$
catalog magnitude. The color of the data point show the Sextractor parameter
ClassStar which gives an estimate of the stellarity of the object. Objects that
have been rejected due to the influence of nearby bright stars are not shown.}
\label{show_objects}
\end{figure}

Now that we have determined a reliable variability estimator and rejected
objects influenced by diffraction spikes, saturated objects, stars and
supernovae, we can go ahead and asses the properties of the final variable AGN
sample.

Figure \ref{show_objects} shows the flux-corrected $C$ for a 0.36\arcsec
aperture versus the object magnitude for the final sample. Variability is
detected down to magnitudes as faint as 25.5, with the faintest object being
clearly detected with a significance of 99.99\%. All of the variable sources
brighter than 18th magnitude are galaxies. At fainter magnitudes, both
point-like and extended objects show in our variability sample, the
most variable objects tend to be point-like.

To assess if the signal to noise limit used in our data is too strict, the
detection rate for the faintest sources is derived. For magnitudes $>$ 25, the
detection rate is close to the expected false positive rate. This indicates
that a lower signal to noise limit would only result in more false positives
and not more 'real' detections.

With 11931 objects with five epoch detections in our catalog, false positive
detections might pose a serious problem. To get an estimate for the expected
false positives contamination, we conservatively assume that all objects are
non-variable and calculate the number of false positives at a given significance.

For a significance of 95\% we expect as many as 596 false positives. Even at a
significance at 99\%, the number of expected false positives is still very high
(119). Only for a significance of 99.9\% does the number of expected false
positives drop to 12. At a significance level of 99.99\%, we expect only a
single false positive, making a catalog with such a strict selection criteria
'clean' from false positives.

\begin{figure}
\includegraphics[width=8cm]{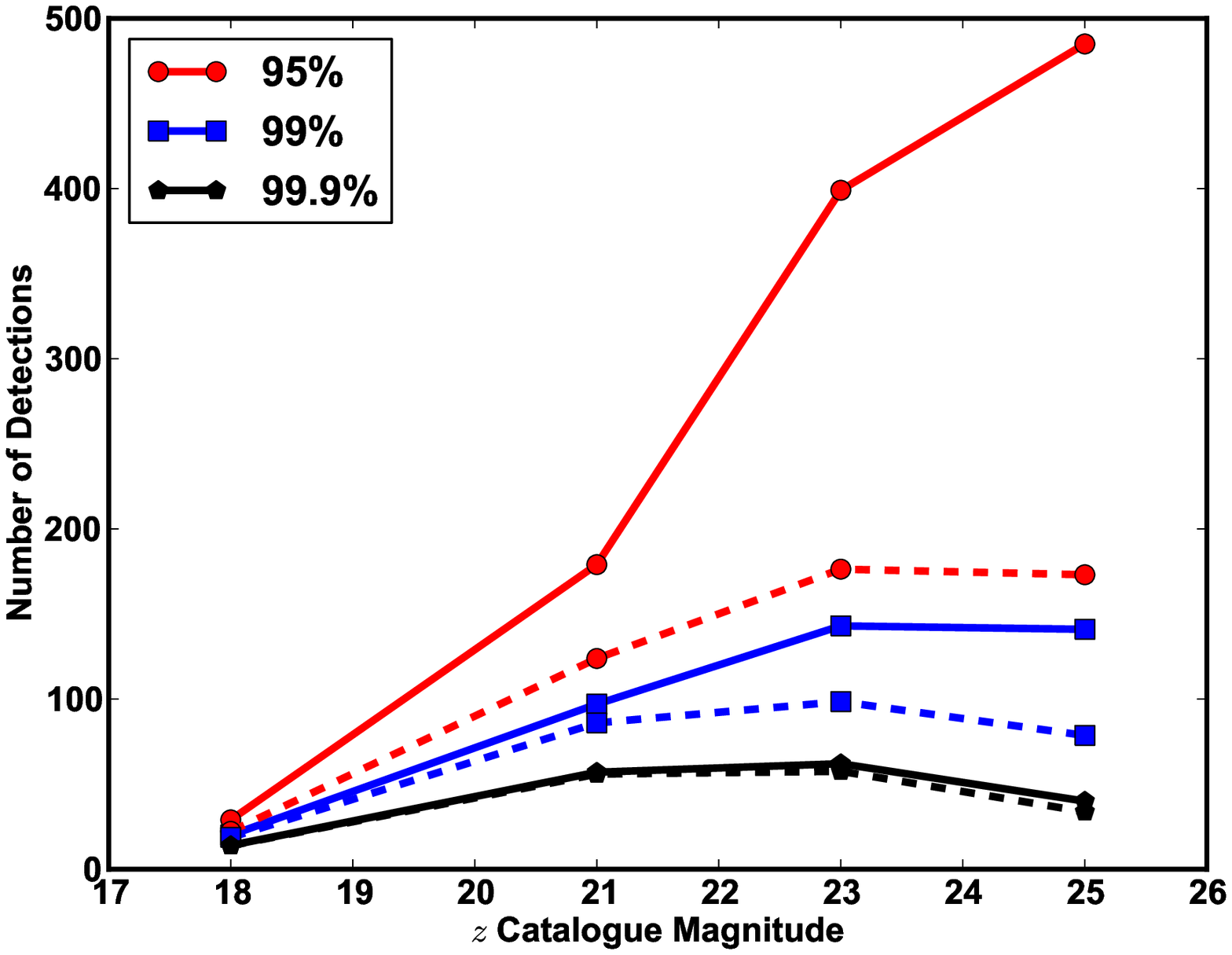}
\caption{Raw (solid line) and false positive corrected (dashed line) number of
detections for different magnitude bins and significance levels.}
\label{detects}
\end{figure}

To see how this affects real data, we show the number of raw and false positive
corrected detections in a number of magnitude bins (Fig. \ref{detects}).
False positive corrections are applied by assuming that all objects are
non-variable and subtracting the expected number of false positives from the
number of detections. We see that false positives pose a very serious problem at
all significance levels lower than 99.9\%. Thus, this very strict limit should be
used when determining the variable object catalog.

However, lower significance levels can be used when trying to estimate the number
of variable objects in the field for statistical arguments. If one is not
interested in knowing the individual objects that show variability but just the
number of variable object in a given subsample, slack significance levels can
be useful as they reduce the number of false negatives. This becomes evident
when looking at Fig. \ref{detects}, there is a big number of false negatives at
a 99.9\% significance level.

To estimate the true number of variables, we use a relaxed significance level of
95\%, this gives us 1072 'variable' objects, given the expected number of false
positive (596) at this significance, we are left with 476 'true' variable
objects, from those, we subtract the 18 objects found to show false variability
due to spikes from bright objects. According to our findings, about 3.8\% of all
objects are variable. This however, does not account for contamination due to
stars. In our final 99.9\% significance catalog with 155 entries, 15 objects are
stars. More objects might turn out to be stars, but could not be
identified as such. Therefore, at least 10\% of all variable objects
identified in this study turn out to be stars.Assuming that
10\% of all variable objects are stars, we derive that about 3.4\% of all
objects in our sample are variable AGN. (Note that the precentage of
stars might depend on the magnitude and therefore our assumption of a fixed
rate of stars might not be correct.)

For the catalog, a significance level of 99.9\% will be used. The final sample
of variables therefore contains 155 objects, with catalog $z$ magnitudes
between 16.45 and 25.5. As mentioned above, the conservatively estimated number
of false positives is 12, resolution in an expected catalog contamination of
about 7.7\%. Out of these 155 variable objects, 15 are identified as stars and
one object is a supernova, leaving 139 variable AGN.

Additionally, a 'clean' sub-catalog with a selection criterion
of 99.99\% is provided. This catalog contains 93 objects and has only one
expected false positive entry. This sub-catalog contains 10 stars and one
supernovae, leaving 82 AGN.

Both final catalogs are shown in Table \ref{cat}. A flag in the table
indicated if the objects belongs to the 'normal' (99.9\% significance) or 'clean'
(99.99\% significance) catalog. The entire catalog will be made available at
Vizier\footnote{http://vizier.u-strasbg.fr/viz-bin/VizieR}.

\subsection{Variability strength}

Additionally, we derive the intrinsic variability for the variability selected
sample. The observed variability can be written as:

\begin{equation}
\sigma_{observed} = \sqrt{\langle\sigma\rangle^{2} + \sigma_{intrinsic}^{2}}
\end{equation}
Where $\sigma$ is the measurement error. This gives the percentage variability
$V$:

\begin{equation}
V = \dfrac{\sigma_{intrinsic}}{\langle flux \rangle} =
\dfrac{\sqrt{\sigma_{observed}^{2} - \langle\sigma\rangle^{2}}}{\langle flux \rangle}
\end{equation}

\begin{figure}
\includegraphics[width=8cm]{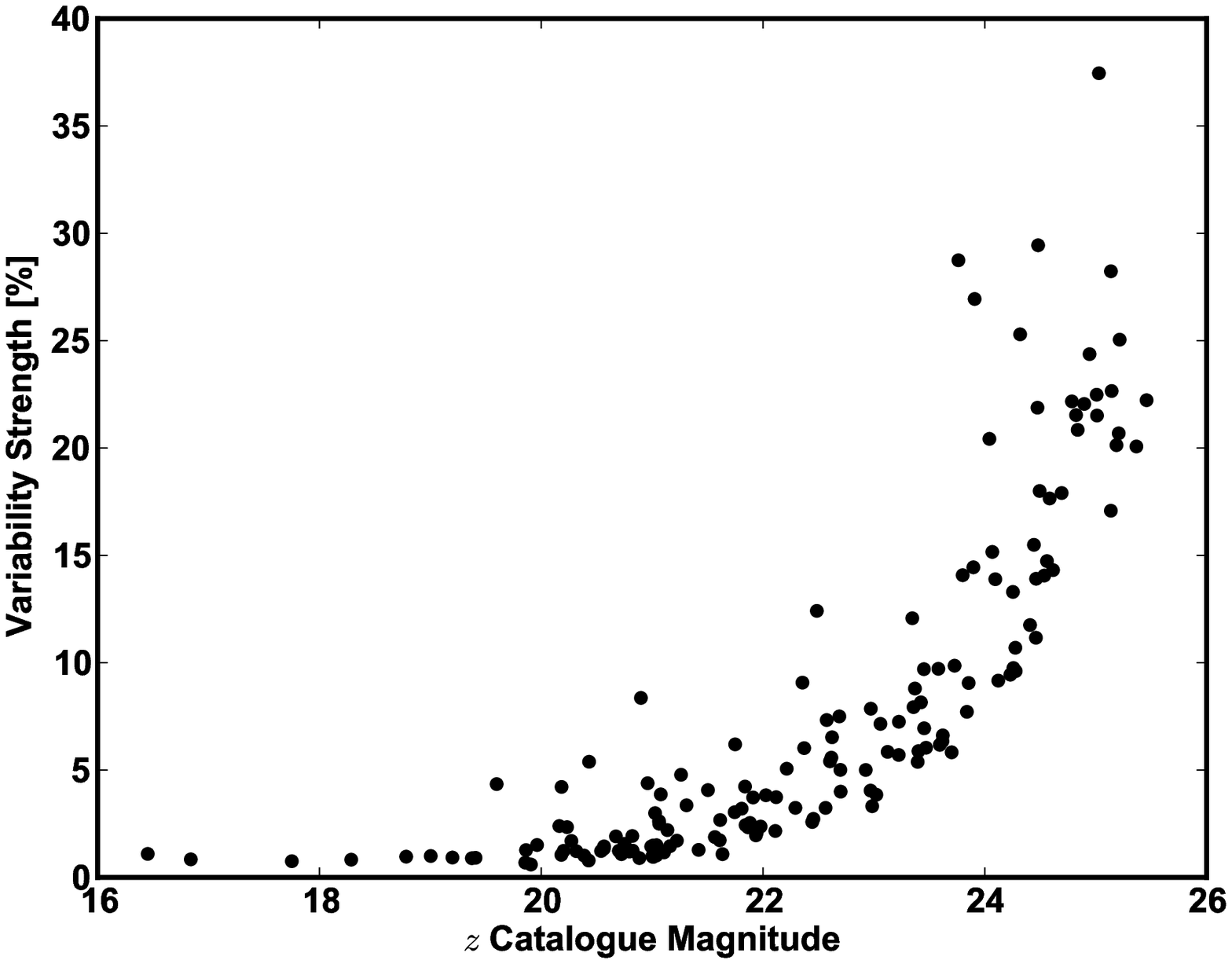}
\caption{Variability Strength $V$ in per cent over catalog magnitude for the
final sample of 156 variable sources.}
\label{V}
\end{figure}

The percentage variability $V$ for all objects in our variable sample is shown in
Figure \ref{V}. Naturally, there is a lower envelope to the variability strength
that is rising to lower luminosity objects, caused by the fact that the signal to
noise ratio is worse for lower luminosities, making it impossible to detect very
subtle variability. $V$ for all variable objects is also shown in the
catalog table (Table \ref{cat}).

To asses the detection limit, we fit a lower envelope to the data. The data is
well fit by the following third order polynomial:

\begin{equation}
V = 0.036 \times mag^{3} -1.504 \times mag^{2} + 15.993 \times mag
-1.644
\end{equation}

This equation holds for magnitudes $>$ 21, for lower magnitudes, a lower
detection limit of about 1\% variability strength is found.

\section{Discussion}

\subsection{Properties of variability selected AGN candidates}

All variability selected objects are matched to the GOODS spectroscopic data
\citep{vanzella_great_2008,popesso_great_2009}. Redshifts, flags and absolute
magnitudes of those objects are included in Table \ref{cat}. 28 out of 155
objects can be matched to the spectroscopic data. Out of those 28 objects, 5 have
been identified as stars. For 3 objects, no spectroscopic redshift could be
derived. The remaining 20 objects have redshifts between 0.045 and 3.7. Amongst
those are five broad-lined AGN, with redshifts of 0.74,0.84,1.23,1.61 and 2.80.
With absolute rest-frame $z$ band magnitudes ranging between -22.57 and
-24.31, all spectroscopically identified AGN in our sample are rather faint. One spectroscopically identified
object has an inferred absolute magnitude of -12.8, indicating that the redshift
identification might be faulty. However the redshift estimation is based on a
single line and might therefore not be correct. Practically all sources show
extremely strong line emission, indicating either high star formation rates or AGN activity.

All variability selected objects are visually inspected. 41 objects are
unresolved, out of those 15 are identified as stars, leaving 26 unresolved AGN
candidates. 18 objects have a dominant core and show faint extended emission.
Nine objects show clear signs of interaction, either in the merger stage or
showing tidal tails. 16 objects are elliptical galaxies and four show clear
disk structure. 41 are resolved, but no morphology can be determined. Those are
mostly faint objects of small size. One object (J033241.87-274651.1) has been
identified by \cite{straughn_tracing_2006} as a tadpole galaxies. Those interesting objects
are believed to be galaxies in an early stage of merging
\citep{straughn_tracing_2006}. However, this particular object is a rather
extreme example of this object class as it shows much less substructure than most
tadpole galaxies. It looks similar to a strong lensed galaxy. However, there are
no clusters nearby.

Finally, seven objects show complex structures. Those galaxies either have
multiple cores or are clearly extended with no clear center. Note that objects
with multiple centers might be wrongly identified as variables due to PSF
changes. This can happen even in cases in which the aperture is big enough to
avoid false variability due to PSF changes for single-center objects. However,
one of those complex sources (J033228.30-274403.6) is detected in X-rays. This
object is also a B-band drop-out, indicating a redshifts around 2-3. Given
that hot cluster gas emission is not detectable at such high redshifts, this
indicates that there is indeed an AGN in this object! Therefore, complex objects
will be included in the final catalog.

\subsection{Comparison with other variable AGN catalogs in the GOODS field}

Now that we have presented our variability-selected AGN catalog and the
estimated occurrence of variability, we would like to compare our sample to other
samples derived in a similar way in the same field.

\cite{sarajedini_v-band_2003} studied HDF data and found 16 variability selected
AGN, none of those is found to be variable in our sample. They performed their
study in the $V$-band, which should pick up different types of variability than
our $z$-band study\citep{di_clemente_variability_1996}. Additionally, they are
covering a time-span of five years while our observations span only about a year.
This will result in higher sensitivity for the detection of more luminous AGN
\citep[e.g.][]{trevese_ensemble_1994,wold_dependence_2007}.
\cite{sarajedini_v-band_2003} used the HDF data, therefore the field is much
smaller than ours but the data is deeper. All in all, their data is very
different from ours, therefore, the missing overlap is somewhat expected.

\cite{cohen_clues_2006} studied $i$-band HUDF data and found 45 'best variable
candidates'. The $i$-band used by \cite{cohen_clues_2006} is only slightly bluer
than the $z$-band used in this study. Therefore, the wavelength dependence of
AGN variability between those bands should be negligible. However, they
covered a time-span of only about four months, opposed to about a year in this
study. This makes their study less sensitive to variability on time
scales of months and longer. The area covered by \cite{cohen_clues_2006} is
much smaller than ours, but their data is significantly deeper. Out of the 45 variable objects identified by
\cite{cohen_clues_2006}, two are found in our catalog as variable. However, the
objects found variable by \cite{cohen_clues_2006} are mostly fainter than 25th
magnitude and the two objects that are found in both catalogs are amongst the
brightest in the Cohen catalog. Given that their data is much deeper, it is
expected that little overlap is found between their and our sample.

\cite{klesman_optical_2007} studied a sample of 112 X-ray and infrared selected
AGN candidates they selected from the GOODS $V$ band five epoch
catalogue. 29 of those objects showed variability. They used the $V$-band data
as AGN variability is found to be stronger at bluer wavelengths \citep[see e.g.][]{di_clemente_variability_1996}. On the other hand, the
$z$-band data used in this study is deeper than the $V$-band data used by
\cite{klesman_optical_2007}. The time coverage for the
\cite{klesman_optical_2007} is practically identical to the one in this study.
Their sample was drawn just from the GOODS South field and was
also restricted to AGN candidates that were preselected, thus also accounting
for the difference in sample size.

Out of the 29 variables from the \cite{klesman_optical_2007} sample, 13 are
detected variable in our study, 12 of those are in our clean catalog, two
objects are too faint to be included in our sample and two objects have been
rejected due to suspiciously high error bars. Two objects have been detected
variable in our study but not by \cite{klesman_optical_2007}. All other objects
are not detected variable in both studies. This is a rather promising overlap.

The differences might be due to several reasons. The different waveband
used might play a role since variability is stronger at shorter wavelengths
\citep{di_clemente_variability_1996}. Additionally, we used a larger aperture
than \cite{klesman_optical_2007}, which we chose in order to reduce PSF-related
variability to less than 0.15 percent. Finally, the variability detection for
our $z$-band selected sample becomes limited beyond about z~25, although a
direct comparison with the $V$-band sample would also need to include
differences in variability amplitude and color terms between the two bands.

\cite{trevese_variability-selected_2008} selected variable AGN from the two year
Southern inTermediate Redshift ESO Supernova Search (STRESS) data. Data were
 taken about every three months over a timespan of two years. Therefore,
\cite{trevese_variability-selected_2008} not only cover double the time span
covered in this study, they also have more data points available, making
it more likely to detect variability. They used $V$ band data taken with the
ground-based ESO/MPI 2.2 m telescope at ESO, La Silla (Chile). With a seeing of
about 1\arcsec, their resolution is about 10 times poorer than that in all
other studies discussed here \citep[][this
study]{sarajedini_v-band_2003,klesman_optical_2007,cohen_clues_2006}. The poorer
resolution makes \cite{trevese_variability-selected_2008} less sensitive to low
luminosity AGN with clearly extended host galaxies. The area covered in their
study is about six times larger than one GOODS Field and therefore about three
times larger than our entire field.

From the 112 variable objects found by \cite{trevese_variability-selected_2008}
only 17 are also in the GOODS footprint. Out of those, three have been detected
variable in our study, six objects did not have detections in all epochs, five
are not variable in our study, two objects were too faint for our signal to noise
cutoff and one object would have been labeled saturated in our study. Given the
big differences between their and our study, such a small overlap is expected.

\section{Summary and Conclusions}

In this paper, we have presented and discussed different statistical methods
that can be used to detect variable sources in large samples. Results have been
applied to the GOODS North and South five epoch $z$-band data.

All three statistical methods tested ($C$,$F$,$\chi^{2}$) show equal powers for
mock data, however, in cases in which the error determinations are
erroneous, $\chi^{2}$ shows the highest power, followed by $C$ and $F$. We have presented a robust
statistical method for detecting variable objects from sparsely sampled data. Our
method makes it possible to control the number of false positives and to test and
correct for possible problems in the error determination.

We present a final catalog of 155 variable objects, selected with a 99.9\%
significance value. Out of those 155 objects, 15 are identified as stars and one
objects is a supernova, leaving 139 variability selected AGN
candidates. The AGN candidates have magnitudes between 16.5
and 25.5 mag in the $z$ band. The catalog has a expected false positive contamination of 12 objects,
corresponding to a contamination of 7.7\% in the entire variable catalog.
Additionally, we present a 'clean catalog' with a significance of 99.99\%. This
catalog contains 93 objects and 82 AGN candidates. It is expected to contain
only a single false positive, corresponding to a contamination rate of about 1\%.
Detection rates at lower significance levels indicate that in total about 3.5\%
of objects are variable AGN. This is higher than the variables rates claimed in
previous studies
\citep{sarajedini_v-band_2003,cohen_clues_2006,klesman_optical_2007}.

20 variability selected AGN have spectroscopic redshifts. The redshifts of those
objects lie between 0.045 and 3.7. Amongst the 20 objects with optical spectra
are five broad-lined AGN, with redshifts between 0.7 and 2.8. With absolute
magnitudes ranging between -22.57 and -24.31, all spectroscopically identified
AGN in our sample are rather faint. This shows that this method is indeed
suitable for detecting low luminosity AGN up to very high redshifts.

Compared to other variability selected catalogs published so far, our
GOODS catalog is larger than the one published by \cite{sarajedini_v-band_2003}
on the Hubble Deep Field which covers a much smaller area, but goes deeper.
Similarly, compared to \cite{cohen_clues_2006} who covered the Hubble Ultra Deep
Field, our sample also covers a larger area with about 3 times more sources, but
is less deep. Since we study both GOODS fields and don't limit our sample to a
preselected AGN catalog, our sample is also larger than that of
\cite{klesman_optical_2007}, and our chosen aperture size reduces our
susceptibility to false positives. Our area covered is smaller than the one in
\cite{trevese_variability-selected_2008}, and they also have a better time
sampling. However, their ground-based data has poorer resolution, making it much
less sensitive to the interesting sample of low-luminosity AGN. Finally, our
study, along with that of \cite{sarajedini_v-band_2003}, are the only ones that
provide estimates for false positive rates.

We have therefore presented a new extensive catalog of variability selected AGN
in the GOODS Fields. Our catalog contains interesting low-luminosity AGN that
cannot be detected using commonly used selection techniques. This makes our
catalog an interesting probe for the poorly understood population of high-redshift
low-luminosity AGN.

Given the formidable multi-wavelength coverage of both GOODS fields and the
availability of both photometric and spectroscopic redshifts, this sample can
give unique insights into the properties of AGN up to high
redshifts and open the window to a sample of previously undetected
low-luminosity high redshift AGN.

The multi-wavelength properties and parent population of our variability selected
AGN sample will be presented in an upcoming paper.

\section*{Acknowledgments} We would like to thank the anonymous referee for
helpful comments. Partial support for this work was provided by NASA through
grant HST-GO-09935.01-A from the Space Telescope Science Institute, which is
operated by the Association of Universities for Research in Astronomy,
Incorporated, under NASA contract NAS5-26555. We acknowledge the help and support
from H. Ferguson in providing advice on running the goodsphot script, which uses
daophot to carry out photometry on the images. We acknowledge M. Paolillo for
helpful discussion, and Vicki Sarajedini for helpful comments and a
thorough reading of the paper.

\bibliographystyle{apj}
\bibliography{variability}

\begin{thebibliography}{63}
\expandafter\ifx\csname natexlab\endcsname\relax\def\natexlab#1{#1}\fi

\bibitem[{Baldwin {et~al.}(1981)Baldwin, Phillips, \&
  Terlevich}]{baldwin_classification_1981}
Baldwin, J.~A., Phillips, M.~M., \& Terlevich, R. 1981, Publications of the
  Astronomical Society of the Pacific, 93, 5

\bibitem[{Becker {et~al.}(1995)Becker, White, \& Helfand}]{becker_first_1995}
Becker, R.~H., White, R.~L., \& Helfand, D.~J. 1995, The Astrophysical Journal,
  450, 559

\bibitem[{Bouwens {et~al.}(2010)Bouwens, Illingworth, Oesch, Stiavelli, van
  Dokkum, Trenti, Magee, Labbé, Franx, Carollo, \&
  Gonzalez}]{bouwens_discovery_2010}
Bouwens, R.~J., {et~al.} 2010, The Astrophysical Journal, 709, L133

\bibitem[{Cohen {et~al.}(2006)Cohen, Ryan, Straughn, Hathi, Windhorst,
  Koekemoer, Pirzkal, Xu, Mobasher, Malhotra, Strolger, \&
  Rhoads}]{cohen_clues_2006}
Cohen, S.~H., {et~al.} 2006, The Astrophysical Journal, 639, 731

\bibitem[{Cristiani {et~al.}(1996)Cristiani, Trentini, Franca, Aretxaga,
  Andreani, Vio, \& Gemmo}]{cristiani_optical_1996}
Cristiani, S., Trentini, S., Franca, F.~L., Aretxaga, I., Andreani, P., Vio,
  R., \& Gemmo, A. 1996, Astronomy and Astrophysics, 306, 395

\bibitem[{de~Diego(2010)}]{de_diego_testing_2010}
de~Diego, J.~A. 2010, 1001.2543

\bibitem[{de~Diego {et~al.}(1998)de~Diego, {Dultzin-Hacyan}, Ramirez, \&
  Benitez}]{de_diego_comparative_1998}
de~Diego, J.~A., {Dultzin-Hacyan}, D., Ramirez, A., \& Benitez, E. 1998,
  Astrophysical Journal, 501, 69

\bibitem[{de~Vries {et~al.}(2005)de~Vries, Becker, White, \&
  Loomis}]{de_vries_structure_2005}
de~Vries, W.~H., Becker, R.~H., White, R.~L., \& Loomis, C. 2005, Astronomical
  Journal, 129, 615

\bibitem[{di~Clemente {et~al.}(1996)di~Clemente, Giallongo, Natali, Trevese, \&
  Vagnetti}]{di_clemente_variability_1996}
di~Clemente, A., Giallongo, E., Natali, G., Trevese, D., \& Vagnetti, F. 1996,
  Astrophysical Journal, 463, 466

\bibitem[{di~Nino {et~al.}(2008)di~Nino, Makidon, Lallo, Sahu, Sirianni, \&
  Casertano}]{di_nino_hst_2008}
di~Nino, D., Makidon, R.~B., Lallo, M., Sahu, K.~C., Sirianni, M., \&
  Casertano, S. 2008, {HST} Focus Variations with Temperature, Tech. rep.

\bibitem[{Dickinson {et~al.}(2003)Dickinson, Giavalisco, \&
  Team}]{dickinson_great_2003}
Dickinson, M., Giavalisco, M., \& Team, G. 2003, 324

\bibitem[{Dunlop \& Peacock(1990)}]{dunlop_redshift_1990}
Dunlop, J.~S., \& Peacock, J.~A. 1990, Monthly Notices of the Royal
  Astronomical Society, 247, 19

\bibitem[{Ferrarese \& Merritt(2000)}]{ferrarese_fundamental_2000}
Ferrarese, L., \& Merritt, D. 2000, The Astrophysical Journal, 539, L9

\bibitem[{Garofalo {et~al.}(2010)Garofalo, Evans, \&
  Sambruna}]{garofalo_evolution_2010}
Garofalo, D., Evans, D.~A., \& Sambruna, R.~M. 2010, 1004.1166

\bibitem[{Gebhardt {et~al.}(2000)Gebhardt, Bender, Bower, Dressler, Faber,
  Filippenko, Green, Grillmair, Ho, Kormendy, Lauer, Magorrian, Pinkney,
  Richstone, \& Tremaine}]{gebhardt_relationship_2000}
Gebhardt, K., {et~al.} 2000, The Astrophysical Journal, 539, L13

\bibitem[{Genzel {et~al.}(1998)Genzel, Lutz, Sturm, Egami, Kunze, Moorwood,
  Rigopoulou, Spoon, Sternberg, {Tacconi-Garman}, Tacconi, \&
  Thatte}]{genzel_what_1998}
Genzel, R., {et~al.} 1998, The Astrophysical Journal, 498, 579

\bibitem[{Giavalisco {et~al.}(2004)Giavalisco, Ferguson, Koekemoer, Dickinson,
  Alexander, Bauer, Bergeron, Biagetti, Brandt, Casertano, Cesarsky,
  Chatzichristou, Conselice, Cristiani, Costa, Dahlen, de~Mello, Eisenhardt,
  Erben, Fall, Fassnacht, Fosbury, Fruchter, Gardner, Grogin, Hook,
  Hornschemeier, Idzi, Jogee, Kretchmer, Laidler, Lee, Livio, Lucas, Madau,
  Mobasher, Moustakas, Nonino, Padovani, Papovich, Park, Ravindranath, Renzini,
  Richardson, Riess, Rosati, Schirmer, Schreier, Somerville, Spinrad, Stern,
  Stiavelli, Strolger, Urry, Vandame, Williams, \&
  Wolf}]{giavalisco_great_2004}
Giavalisco, M., {et~al.} 2004, Astrophysical Journal, 600, L93

\bibitem[{Graham {et~al.}(2001)Graham, Erwin, Caon, \&
  Trujillo}]{graham_correlation_2001}
Graham, A.~W., Erwin, P., Caon, N., \& Trujillo, I. 2001, The Astrophysical
  Journal, 563, L11

\bibitem[{Groenewegen {et~al.}(2002)Groenewegen, Girardi, Hatziminaoglou,
  Benoist, Olsen, da~Costa, Arnouts, Madejsky, Mignani, Rité, Sikkema,
  Slijkhuis, \& Vandame}]{groenewegen_eso_2002}
Groenewegen, M. A.~T., {et~al.} 2002, Astronomy and Astrophysics, 392, 741

\bibitem[{Hatziminaoglou {et~al.}(2002)Hatziminaoglou, Groenewegen, da~Costa,
  Arnouts, Benoist, Madejsky, Mignani, Olsen, Rité, Schirmer, Slijkhuis, \&
  Vandame}]{hatziminaoglou_eso_2002}
Hatziminaoglou, E., {et~al.} 2002, Astronomy and Astrophysics, 384, 81

\bibitem[{Hornschemeier {et~al.}(2001)Hornschemeier, Brandt, Garmire,
  Schneider, Barger, Broos, Cowie, Townsley, Bautz, Burrows, Chartas,
  Feigelson, Griffiths, Lumb, Nousek, Ramsey, \&
  Sargent}]{hornschemeier_chandra_2001}
Hornschemeier, A.~E., {et~al.} 2001, The Astrophysical Journal, 554, 742

\bibitem[{Hutchings(2003)}]{hutchings_host_2003}
Hutchings, J.~B. 2003, Astronomical Journal, 125, 1053

\bibitem[{Jahnke {et~al.}(2004)Jahnke, S�nchez, Wisotzki, Barden, Beckwith,
  Bell, Borch, Caldwell, H�ussler, Heymans, Jogee, {McIntosh}, Meisenheimer,
  Peng, Rix, Somerville, \& Wolf}]{jahnke_ultraviolet_2004}
Jahnke, K., {et~al.} 2004, Astrophysical Journal, 614, 568

\bibitem[{Jansen {et~al.}(2001)Jansen, Lumb, Altieri, Clavel, Ehle, Erd,
  Gabriel, Guainazzi, Gondoin, Much, Munoz, Santos, Schartel, Texier, \&
  Vacanti}]{jansen_xmm-newton_2001}
Jansen, F., {et~al.} 2001, Astronomy and Astrophysics, 365, L1

\bibitem[{Jiang {et~al.}(2010)Jiang, Fan, Brandt, Carilli, Egami, Hines, Kurk,
  Richards, Shen, Strauss, Vestergaard, \& Walter}]{jiang_dust-free_2010}
Jiang, L., {et~al.} 2010, 1003.3432

\bibitem[{Klesman \& Sarajedini(2007)}]{klesman_optical_2007}
Klesman, A., \& Sarajedini, V. 2007, Astrophysical Journal, 665, 225

\bibitem[{Koekemoer {et~al.}(2002)Koekemoer, Fruchter, Hook, \&
  Hack}]{koekemoer_multidrizzle:integrated_2002}
Koekemoer, A.~M., Fruchter, A.~S., Hook, R.~N., \& Hack, W. 2002,HST Calibration Workshop (eds. S. Arribas, A. Koekemoer \&
B. Whitmore, STScI: Baltimore), 337

\bibitem[{Krist(1995)}]{krist_simulation_1995}
Krist, J. 1995, in , 349

\bibitem[{Madau {et~al.}(1998)Madau, Pozzetti, \& Dickinson}]{madau_star_1998}
Madau, P., Pozzetti, L., \& Dickinson, M. 1998, The Astrophysical Journal, 498,
  106

\bibitem[{Marconi \& Hunt(2003)}]{marconi_relation_2003}
Marconi, A., \& Hunt, L.~K. 2003, Astrophysical Journal, 589, L21

\bibitem[{Markarian(1967)}]{markarian_galaxies_1967}
Markarian, B.~E. 1967, Astrofizika, 3, 55

\bibitem[{Mendez \& Guzman(1998)}]{mendez_starcounts_1998}
Mendez, R.~A., \& Guzman, R. 1998, Astronomy and Astrophysics, 333, 106

\bibitem[{Paolillo {et~al.}(2004)Paolillo, Schreier, Giacconi, Koekemoer, \&
  Grogin}]{paolillo_prevalence_2004}
Paolillo, M., Schreier, E.~J., Giacconi, R., Koekemoer, A.~M., \& Grogin, N.~A.
  2004, Astrophysical Journal, 611, 93

\bibitem[{Popesso {et~al.}(2009)Popesso, Dickinson, Nonino, Vanzella, Daddi,
  Fosbury, Kuntschner, Mainieri, Cristiani, Cesarsky, Giavalisco, Renzini, \&
  Team}]{popesso_great_2009}
Popesso, P., {et~al.} 2009, Astronomy and Astrophysics, 494, 443

\bibitem[{Reddy {et~al.}(2006)Reddy, Steidel, Erb, Shapley, \&
  Pettini}]{reddy_spectroscopic_2006}
Reddy, N.~A., Steidel, C.~C., Erb, D.~K., Shapley, A.~E., \& Pettini, M. 2006,
  Astrophysical Journal, 653, 1004

\bibitem[{Richards {et~al.}(2002)Richards, Fan, Newberg, Strauss, Berk,
  Schneider, Yanny, Boucher, Burles, Frieman, Gunn, Hall, Željko Ivezić,
  Kent, Loveday, Lupton, Rockosi, Schlegel, Stoughton, {SubbaRao}, \&
  York}]{richards_spectroscopic_2002}
Richards, G.~T., {et~al.} 2002, The Astronomical Journal, 123, 2945

\bibitem[{Riess {et~al.}(2004)Riess, Strolger, Tonry, Casertano, Ferguson,
  Mobasher, Challis, Filippenko, Jha, Li, Chornock, Kirshner, Leibundgut,
  Dickinson, Livio, Giavalisco, Steidel, Benítez, \&
  Tsvetanov}]{riess_type_2004}
Riess, A.~G., {et~al.} 2004, The Astrophysical Journal, 607, 665

\bibitem[{Sanders {et~al.}(1989)Sanders, Phinney, Neugebauer, Soifer, \&
  Matthews}]{sanders_continuum_1989}
Sanders, D.~B., Phinney, E.~S., Neugebauer, G., Soifer, B.~T., \& Matthews, K.
  1989, The Astrophysical Journal, 347, 29

\bibitem[{Santini {et~al.}(2009)Santini, Fontana, Grazian, Salimbeni, Fiore,
  Fontanot, Boutsia, Castellano, Cristiani, de~Santis, Gallozzi, Giallongo,
  Menci, Nonino, Paris, Pentericci, \& Vanzella}]{santini_star_2009}
Santini, P., {et~al.} 2009, Astronomy and Astrophysics, 504, 751

\bibitem[{Sarajedini {et~al.}(2003)Sarajedini, Gilliland, \&
  Kasm}]{sarajedini_v-band_2003}
Sarajedini, V.~L., Gilliland, R.~L., \& Kasm, C. 2003, The Astrophysical
  Journal, 599, 173

\bibitem[{Schmidt \& Green(1983)}]{schmidt_quasar_1983}
Schmidt, M., \& Green, R.~F. 1983, The Astrophysical Journal, 269, 352

\bibitem[{Schramm {et~al.}(2008)Schramm, Wisotzki, \&
  Jahnke}]{schramm_host_2008}
Schramm, M., Wisotzki, L., \& Jahnke, K. 2008, Astronomy and Astrophysics, 478,
  311

\bibitem[{Shemmer {et~al.}(2009)Shemmer, Brandt, Anderson, {Diamond-Stanic},
  Fan, Richards, Schneider, \& Strauss}]{shemmer_x-ray_2009}
Shemmer, O., Brandt, W.~N., Anderson, S.~F., {Diamond-Stanic}, A.~M., Fan, X.,
  Richards, G.~T., Schneider, D.~P., \& Strauss, M.~A. 2009, 0902.1366

\bibitem[{Sirianni {et~al.}(2005)Sirianni, Jee, Benítez, Blakeslee, Martel,
  Meurer, Clampin, Marchi, Ford, Gilliland, Hartig, Illingworth, Mack, \&
  {McCann}}]{sirianni_photometric_2005}
Sirianni, M., {et~al.} 2005, Publications of the Astronomical Society of the
  Pacific, 117, 1049

\bibitem[{Smith \& Wright(1980)}]{smith_radio_1980}
Smith, M.~G., \& Wright, A.~E. 1980, Monthly Notices of the Royal Astronomical
  Society, 191, 871

\bibitem[{Stocke(2001)}]{stocke_hidden_2001}
Stocke, J.~T. 2001, in , 184

\bibitem[{Straughn {et~al.}(2006)Straughn, Cohen, Ryan, Hathi, Windhorst, \&
  Jansen}]{straughn_tracing_2006}
Straughn, A.~N., Cohen, S.~H., Ryan, R.~E., Hathi, N.~P., Windhorst, R.~A., \&
  Jansen, R.~A. 2006, The Astrophysical Journal, 639, 724

\bibitem[{Szokoly {et~al.}(2004)Szokoly, Bergeron, Hasinger, Lehmann, Kewley,
  Mainieri, Nonino, Rosati, Giacconi, Gilli, Gilmozzi, Norman, Romaniello,
  Schreier, Tozzi, Wang, Zheng, \& Zirm}]{szokoly_chandra_2004}
Szokoly, G.~P., {et~al.} 2004, The Astrophysical Journal Supplement Series,
  155, 271

\bibitem[{Taylor {et~al.}(2009)Taylor, Franx, van Dokkum, Quadri, Gawiser,
  Bell, Barrientos, Blanc, Castander, Damen, {Gonzalez-Perez}, Hall, Herrera,
  Hildebrandt, Kriek, Labbé, Lira, Maza, Rudnick, Treister, Urry, Willis, \&
  Wuyts}]{taylor_public_2009}
Taylor, E.~N., {et~al.} 2009, The Astrophysical Journal Supplement Series, 183,
  295

\bibitem[{Timmer \& Koenig(1995)}]{timmer_generating_1995}
Timmer, J., \& Koenig, M. 1995, Astronomy and Astrophysics, 300, 707

\bibitem[{Trevese {et~al.}(2008)Trevese, Boutsia, Vagnetti, Cappellaro, \&
  Puccetti}]{trevese_variability-selected_2008}
Trevese, D., Boutsia, K., Vagnetti, F., Cappellaro, E., \& Puccetti, S. 2008,
  Astronomy and Astrophysics, 488, 73

\bibitem[{Trevese {et~al.}(1994)Trevese, Kron, Majewski, Bershady, \&
  Koo}]{trevese_ensemble_1994}
Trevese, D., Kron, R.~G., Majewski, S.~R., Bershady, M.~A., \& Koo, D.~C. 1994,
  The Astrophysical Journal, 433, 494

\bibitem[{Ulrich {et~al.}(1997)Ulrich, Maraschi, \&
  Urry}]{ulrich_variability_1997}
Ulrich, M., Maraschi, L., \& Urry, C.~M. 1997, Annual Review of Astronomy and
  Astrophysics, 35, 445

\bibitem[{Urry \& Padovani(1995)}]{urry_unified_1995}
Urry, C.~M., \& Padovani, P. 1995, Publications of the Astronomical Society of
  the Pacific, 107, 803

\bibitem[{Vanzella {et~al.}(2008)Vanzella, Cristiani, Dickinson, Giavalisco,
  Kuntschner, Haase, Nonino, Rosati, Cesarsky, Ferguson, Fosbury, Grazian,
  Moustakas, Rettura, Popesso, Renzini, Stern, \& Team}]{vanzella_great_2008}
Vanzella, E., {et~al.} 2008, Astronomy and Astrophysics, 478, 83

\bibitem[{Vaughan {et~al.}(2003)Vaughan, Edelson, Warwick, \&
  Uttley}]{vaughan_characterizingvariability_2003}
Vaughan, S., Edelson, R., Warwick, R.~S., \& Uttley, P. 2003, Monthly Notices
  of the Royal Astronomical Society, 345, 1271

\bibitem[{Veilleux \& Osterbrock(1987)}]{veilleux_spectral_1987}
Veilleux, S., \& Osterbrock, D.~E. 1987, The Astrophysical Journal Supplement
  Series, 63, 295

\bibitem[{Villforth {et~al.}(2008)Villforth, Heidt, \&
  Nilsson}]{villforth_quasar_2008}
Villforth, C., Heidt, J., \& Nilsson, K. 2008, Astronomy and Astrophysics, 488,
  133

\bibitem[{Villforth {et~al.}(2009)Villforth, Nilsson, Østensen, Heidt, Niemi,
  \& Pforr}]{villforth_intranight_2009}
Villforth, C., Nilsson, K., Østensen, R., Heidt, J., Niemi, S., \& Pforr, J.
  2009, Monthly Notices of the Royal Astronomical Society, 397, 1893

\bibitem[{Warren {et~al.}(1991)Warren, Hewett, \& Osmer}]{warren_quasar_1991}
Warren, S.~J., Hewett, P.~C., \& Osmer, P.~S. 1991, in , 139--148

\bibitem[{Weisskopf {et~al.}(2000)Weisskopf, Tananbaum, Speybroeck, \&
  {O'Dell}}]{weisskopf_chandra_2000}
Weisskopf, M.~C., Tananbaum, H.~D., Speybroeck, L. P.~V., \& {O'Dell}, S.~L.
  2000, in , 2--16

\bibitem[{Wold {et~al.}(2007)Wold, Brotherton, \& Shang}]{wold_dependence_2007}
Wold, M., Brotherton, M.~S., \& Shang, Z. 2007, Monthly Notices of the Royal
  Astronomical Society, 375, 989

\bibitem[{Wolf {et~al.}(2008)Wolf, Hildebrandt, Taylor, \&
  Meisenheimer}]{wolf_calibration_2008}
Wolf, C., Hildebrandt, H., Taylor, E.~N., \& Meisenheimer, K. 2008, Astronomy
  and Astrophysics, 492, 933

\end{thebibliography}

\appendix

\section{Results of statistical simulations}

\begin{deluxetable}{cc|cc|cc|cc}
\tabletypesize{\scriptsize}
\tablecaption{Detection power (in per cent) for different mock light curves at a
significance of 99.9 per cent with 'erroneous' errors. Variability
strength $V$ is identical to the one used in Table \ref{power} and defined in
the text. Results are given for an error in the error measurement of 5 and 25 per cent.
\label{power_err}.}
\tablehead{
\colhead{} &
\colhead{Signalstrength} &
\colhead{Slope 5\%} &
\colhead{Slope 25\%} &
\colhead{Burst 5\%} &
\colhead{Burst 25\%} &
\colhead{Noise 5\%} &
\colhead{Noise 25\%}
%& $V$ & Slope & Slope & Burst & Burst & Noise & Noise \\
%&  & 5\% & 25\% & 5\% & 25\% & 5\% & 25\% \\
}
\startdata
$F$ & 0.5 & 0.13 & 0.22 & 0.15 & 0.25 & 0.00 & 0.00 \\
$C$ & 0.5 & 0.14 & 0.26 & 0.15 & 0.34 & 0.00 & 0.00 \\
$\chi^{2}$ & 0.5 & 0.14 & 1.05 & 0.17 & 1.18 & 0.00 & 0.01 \\
$F$ & 1.0 & 0.20 & 0.34 & 0.37 & 0.56 & 0.11 & 0.18 \\
$C$ & 1.0 & 0.20 & 0.43 & 0.37 & 0.69 & 0.11 & 0.23 \\
$\chi^{2}$ & 1.0 & 0.22 & 1.32 & 0.40 & 1.92 & 0.12 & 0.97 \\
$F$ & 1.5 & 0.40 & 0.59 & 0.98 & 1.21 & 8.72 & 8.75 \\
$C$ & 1.5 & 0.41 & 0.76 & 0.99 & 1.47 & 8.80 & 9.68 \\
$\chi^{2}$ & 1.5 & 0.45 & 2.03 & 1.04 & 3.49 & 8.94 & 13.52 \\
$F$ & 2.0 & 0.80 & 0.98 & 2.43 & 2.92 & 32.92 & 32.17 \\
$C$ & 2.0 & 0.81 & 1.20 & 2.47 & 3.43 & 33.07 & 33.80 \\
$\chi^{2}$ & 2.0 & 0.87 & 2.98 & 2.61 & 6.50 & 33.31 & 38.29 \\
$F$ & 2.5 & 1.60 & 1.95 & 5.71 & 6.29 & 56.23 & 54.96 \\
$C$ & 2.5 & 1.62 & 2.33 & 5.79 & 7.23 & 56.36 & 56.57 \\
$\chi^{2}$ & 2.5 & 1.70 & 4.76 & 5.99 & 11.62 & 56.50 & 59.90 \\
$F$ & 3.0 & 3.05 & 3.55 & 12.16 & 12.57 & 72.62 & 71.01 \\
$C$ & 3.0 & 3.10 & 4.17 & 12.29 & 14.14 & 72.73 & 72.22 \\
$\chi^{2}$ & 3.0 & 3.20 & 7.47 & 12.57 & 19.65 & 72.82 & 74.52 \\
$F$ & 3.5 & 5.56 & 6.19 & 22.90 & 22.13 & 82.45 & 81.42 \\
$C$ & 3.5 & 5.63 & 7.08 & 23.10 & 24.23 & 82.53 & 82.31 \\
$\chi^{2}$ & 3.5 & 5.77 & 11.36 & 23.34 & 30.04 & 82.59 & 83.83 \\
$F$ & 4.0 & 9.50 & 9.74 & 36.92 & 35.31 & 88.55 & 87.62 \\
$C$ & 4.0 & 9.62 & 11.11 & 37.19 & 37.92 & 88.59 & 88.24 \\
$\chi^{2}$ & 4.0 & 9.88 & 16.40 & 37.45 & 43.16 & 88.65 & 89.31 \\
$F$ & 4.5 & 15.58 & 15.65 & 53.77 & 50.89 & 92.46 & 91.50 \\
$C$ & 4.5 & 15.73 & 17.30 & 54.06 & 53.67 & 92.50 & 91.94 \\
$\chi^{2}$ & 4.5 & 16.00 & 23.46 & 54.14 & 57.53 & 92.51 & 92.64 \\
$F$ & 5.0 & 23.68 & 23.12 & 69.73 & 66.09 & 94.58 & 94.18 \\
$C$ & 5.0 & 23.86 & 25.31 & 69.96 & 68.56 & 94.60 & 94.50 \\
$\chi^{2}$ & 5.0 & 24.18 & 31.94 & 69.95 & 70.40 & 94.66 & 94.99 \\
$F$ & 5.5 & 33.49 & 31.93 & 82.58 & 79.40 & 96.24 & 95.81 \\
$C$ & 5.5 & 33.73 & 34.54 & 82.75 & 81.21 & 96.26 & 96.04 \\
$\chi^{2}$ & 5.5 & 34.25 & 41.10 & 82.72 & 81.52 & 96.29 & 96.39 \\
$F$ & 6.0 & 45.00 & 42.66 & 91.52 & 88.97 & 97.27 & 97.02 \\
$C$ & 6.0 & 45.26 & 45.37 & 91.61 & 90.22 & 97.28 & 97.19 \\
$\chi^{2}$ & 6.0 & 45.61 & 51.54 & 91.48 & 89.66 & 97.28 & 97.44 \\
$F$ & 6.5 & 56.66 & 53.90 & 96.50 & 94.88 & 97.96 & 97.79 \\
$C$ & 6.5 & 56.93 & 56.63 & 96.54 & 95.57 & 97.97 & 97.90 \\
$\chi^{2}$ & 6.5 & 57.28 & 61.89 & 96.48 & 94.89 & 97.96 & 98.07 \\
$F$ & 7.0 & 68.09 & 64.65 & 98.72 & 97.92 & 98.39 & 98.33 \\
$C$ & 7.0 & 68.33 & 67.20 & 98.74 & 98.21 & 98.40 & 98.42 \\
$\chi^{2}$ & 7.0 & 68.55 & 71.43 & 98.72 & 97.72 & 98.40 & 98.54 \\
$F$ & 7.5 & 78.02 & 74.83 & 99.66 & 99.27 & 98.77 & 98.69 \\
$C$ & 7.5 & 78.20 & 76.94 & 99.67 & 99.39 & 98.77 & 98.78 \\
$\chi^{2}$ & 7.5 & 78.31 & 79.91 & 99.64 & 99.10 & 98.77 & 98.90 \\
$F$ & 8.0 & 85.96 & 82.80 & 99.92 & 99.78 & 99.08 & 98.94 \\
$C$ & 8.0 & 86.11 & 84.43 & 99.93 & 99.82 & 99.08 & 99.01 \\
$\chi^{2}$ & 8.0 & 86.16 & 86.35 & 99.92 & 99.67 & 99.08 & 99.09 \\
$F$ & 8.5 & 91.73 & 89.12 & 99.98 & 99.95 & 99.25 & 99.18 \\
$C$ & 8.5 & 91.82 & 90.28 & 99.98 & 99.96 & 99.26 & 99.22 \\
$\chi^{2}$ & 8.5 & 91.81 & 91.57 & 99.98 & 99.91 & 99.25 & 99.31 \\
$F$ & 9.0 & 95.46 & 93.57 & 100.00 & 99.99 & 99.40 & 99.35 \\
$C$ & 9.0 & 95.52 & 94.37 & 100.00 & 99.99 & 99.40 & 99.38 \\
$\chi^{2}$ & 9.0 & 95.52 & 94.99 & 100.00 & 99.97 & 99.41 & 99.45 \\
$F$ & 9.5 & 97.70 & 96.39 & 100.00 & 100.00 & 99.48 & 99.44 \\
$C$ & 9.5 & 97.72 & 96.87 & 100.00 & 100.00 & 99.48 & 99.47 \\
$\chi^{2}$ & 9.5 & 97.75 & 97.22 & 100.00 & 100.00 & 99.49 & 99.52 \\
\enddata
\end{deluxetable}

%\clearpage

\section{Variable object catalog}

\begin{deluxetable}{lcccccccccc}
\tabletypesize{\scriptsize}
\tablewidth{0pt}
\tablecaption{First part of variable objects catalog. Columns: ID: IAU object
identification; RA (J2000): Right Ascension (Julian Year 2000); Dec (J2000):
Declination (Julian Year 2000); mag ($z$): Sextractor $z$ band magnitude; CS:
stellarity (1 is point like, 0 is extended); $C$: flux corrected $C$ for 0.36
\arcsec aperture; $V$ (\%): variability strength in percent; Catalog: Normal
(99.9\% insignificance) or Clean (99.99\% significance); Comment: morphology or
classification, references for classifications are given; $z$: spectroscopic
redshift if available and flag (secure (S), likely (L), tentative (T)), 'None' if spectroscopy is available but no redshift
was found. Unless otherwise noted, all spectroscopic redshifts are from
\cite{popesso_great_2009}; $M_{abs}$: absolute magnitude in observed frame $z$
(only for objects with spectroscopic redshift $>$ 0). Objects that have been
detected in other studies are marked by superscripts in the ID column:
\citet{cohen_clues_2006}: C, \citet{klesman_optical_2007}: KS,
\citet{trevese_variability-selected_2008}: T.
\label{cat}}
\tablehead{
\colhead{ID} &
\colhead{RA (J2000)} &
\colhead{Dec (J2000)} &
\colhead{mag ($z$)} &
\colhead{CS} &
\colhead{$C$} &
\colhead{$V$ (\%)} &
\colhead{Catalog} &
\colhead{Comment} &
\colhead{$z$} &
\colhead{$M_{abs}$}
}
\startdata
J033203.00-274213.6 & 53.01248 & -27.7037872 & 25.14 & 0.09 & 2.55 & 28.23 & Clean & Resolved &... & ...\\
J033203.01-274544.7 & 53.0125331 & -27.7624232 & 24.90 & 0.02 & 2.05 & 22.05 & Normal & Resolved &...  & ...\\
J033203.26-274530.3 & 53.013574 & -27.7584257 & 22.70 & 0.03 & 2.15 & 3.99 & Normal & Resolved &...  & ...\\
J033204.41-274635.5 & 53.018367 & -27.7765167 & 21.14 & 0.98 & 3.92 & 2.20 & Clean & Unresolved &...  & ...\\
J033205.11-274317.5 & 53.021302 & -27.7215411 & 21.91 & 0.98 & 4.30 & 3.72 & Clean & Unresolved &...  & ...\\
J033205.40-274429.2 & 53.0224983 & -27.7414518 & 22.97 & 0.03 & 2.47 & 7.86 &Clean & Disk Galaxy &None & ...\\
J033208.68-274508.0 & 53.036184 & -27.7522317 & 22.70 & 0.03 & 1.98 & 5.01
&Normal & Resolved &1.2964$^{L}$ & -22.11\\
J033209.57-274634.9 & 53.0398585&-27.7763722 & 22.97 & 0.03 & 2.03 & 4.04 & Normal & Elliptical &... & ...\\
J033209.58-274241.8 & 53.0399142 & -27.7116218 & 23.39 & 0.03 & 2.05 & 5.38 & Normal & Elliptical &... & ...\\
J033209.80-274308.6 & 53.0408277 & -27.7190613 & 23.12 & 0.03 & 1.93 &
5.85&Normal & Interacting system &2.3021$^{S}$ & -23.20\\
J033210.52-274628.9&53.0438334 & -27.7747007 & 23.85 & 0.03 & 2.10 & 9.06 &Normal & Resolved &... & ...\\
J033210.91-274414.9$^{KS}$ & 53.0454704 & -27.7374846 & 22.37 & 0.93 & 4.69 &6.02 &Clean & Elliptical &1.6082$^{L}$ & -23.01\\
J033211.02-274919.8 & 53.045918 &-27.8221721 & 23.45 & 0.03 & 2.15 & 9.70&Normal & Interacting system &1.9431$^{L}$ & -23.42\\
J033212.16-274408.8&53.0506774 & -27.7357915 & 24.79 & 0.03 & 2.29 & 22.17 & Clean & Resolved &...& ...\\
J033213.21-274715.7 & 53.0550544 & -27.7876915 & 24.09 & 0.03 & 2.28 & 13.89 & Clean & Resolved &...& ...\\
J033213.34-274210.5 & 53.0555659 & -27.7029202 & 23.84 & 0.99 & 2.34 & 7.71 & Clean & Unresolved &...& ...\\
J033215.16-274754.6 & 53.0631513 & -27.7985117 & 22.98 & 0.98 & 2.07 & 3.32 & Normal & Unresolved &...& ...\\
J033215.93-275329.3 & 53.0663591 & -27.8914797 & 19.86 & 0.92 & 2.15 &
0.69&Normal & Star\tablenotemark{e} &0.0000$^{S}$ & ...\\
J033216.34-274851.7 & 53.0681031 &-27.8143745 &21.98 & 0.98 & 2.65 & 2.37 & Clean & Unresolved &...& ...\\
J033216.87-274916.7 & 53.0702817 & -27.8212937 & 23.40 & 0.98 & 2.43 & 5.88 &Clean & Unresolved &0.0000$^{L}$& ...\\
J033217.06-274921.9 & 53.0710682 &-27.8227401 & 19.20 & 0.03 & 2.03 &
0.93&Normal & Interacting system &0.3375$^{S}$ & -22.06\\
%\tablebreak
J033217.14-274303.3$^{KS,T}$&53.0714326 & -27.7175864 & 20.57 & 0.03 & 3.14 &1.44 & Clean & Elliptical &...& ...\\
J033217.72-274703.0 & 53.0738469 & -27.7841607 & 23.59 & 0.03 & 1.99 & 6.18 & Normal & Interacting system &... & ...\\
J033218.24-275241.4$^{KS}$ & 53.0760024 & -27.8781606 & 24.28 & 0.99 & 2.15 &
9.61 &Normal & Unresolved &2.8049$^{S}$ & -22.57\\
J033218.70-275149.3 & 53.0778965 &-27.8637054 & 21.74 & 0.03 & 2.20 &3.04&Clean& Disk galaxy &0.4568$^{S}$ & -20.55\\
J033218.81-274908.5 & 53.0783542 &-27.8190409 & 23.90 & 0.03 & 2.55 & 14.44 & Clean & Complex &... & ...\\
J033218.84-274529.2 & 53.078519 & -27.7581146 & 18.78 & 0.03 & 2.44 & 0.97 & Clean & Elliptical &... & ...\\
J033219.81-275300.9 & 53.0825539 & -27.8835727 & 24.46 & 0.03 & 1.95 & 11.16
&Normal & Resolved &3.7072$^{S}$ & -23.12\\
J033219.86-274110.0 & 53.0827492 &-27.686119 & 23.37 & 0.00 & 2.07 & 8.80 &Normal & Interacting system &None & ...\\
J033220.80-275144.5 &53.0866794 & -27.8623513 & 21.22 & 0.98 & 2.91 & 1.71 & Clean & Star &... & ...\\
J033221.52-274358.7 & 53.0896509 & -27.732984 & 21.00 & 0.97 & 1.95 & 0.99 &
Normal & Star\tablenotemark{j} &... & ...\\
J033222.82-274518.4 & 53.0950956 & -27.7550986 & 23.22 & 0.03 & 1.92 & 5.70 & Normal & Resolved &... & ...\\
J033223.53-274707.5 & 53.0980434 & -27.785425 & 23.23 & 0.03 & 2.00 & 7.25 & Normal & Interacting system &... & ...\\
J033224.23-274129.5 & 53.1009433 & -27.691518 & 25.01 & 0.00 & 1.96 & 21.51 & Normal & Complex &... & ...\\
J033224.54-274010.4 & 53.1022685 & -27.6695645 & 21.88 & 0.03 & 1.95 & 2.54 & Normal & Interacting system &... & ...\\
J033224.80-274617.9 & 53.1033449 & -27.7716431 & 23.06 & 0.03 & 2.69 & 7.15 & Clean & Resolved &...& ...\\
J033225.10-274403.2 & 53.1045636 & -27.7342138 & 17.75 & 0.03 & 2.73 & 0.75 & Clean & Elliptical &... & ...\\
J033225.15-274933.3 & 53.1048118 & -27.8259053 & 21.61 & 0.99 & 2.56 & 1.73
&Clean & Star\tablenotemark{e} &0.0000$^{S}$ & ...\\
J033225.99-274142.9 & 53.1082952 & -27.6952603& 24.54 & 0.03 & 1.95 & 14.06&Normal & Resolved &0.0459$^{S}$ & -12.18\\
J033226.40-275532.4 & 53.109991 &-27.9256529 & 24.32 & 0.00 & 2.21 & 25.29 & Clean & Complex &... & ...\\
J033226.49-274035.5$^{KS,T}$ & 53.1103938 & -27.6765399 & 19.60 & 0.93 & 13.45 &
4.35&Clean & Core + spiral structure &0.5404$^{L}$& -22.87\\
J033227.01-274105.0$^{KS}$&53.1125287 & -27.6847238 & 19.00 & 0.93 & 3.97 & 1.00
&Clean & Core + extended emission &0.7423$^{S}$& -24.31\\
J033227.18-274416.5 &53.113269 & -27.73791 & 19.37 & 0.03 & 2.26 & 0.89 & Clean & Elliptical &... & ...\\
J033227.51-275612.4 & 53.114644 & -27.9367764 & 24.04 & 0.03 & 4.18 & 20.42
&Clean & Resolved &0.6630$^{S}$ & -18.97\\
J033227.86-275335.6 & 53.1160832&-27.8932186 & 21.84 & 0.98 & 5.00 & 2.44 & Clean & Star\tablenotemark{a} &...& ...\\
J033227.87-275335.9 & 53.1161233 &-27.8933035 & 21.63 & 0.99 & 2.30 & 1.08 &Clean & Star\tablenotemark{a} &... & ...\\
J033228.30-274403.6 & 53.1179209 & -27.7343234 & 23.76 & 0.00 & 2.25 & 28.74 & Clean & Complex &... & ...\\
J033228.45-274203.8 & 53.118527 & -27.7010451 & 24.23 & 0.99 & 2.08 & 9.44 & Normal & Unresolved &... & ...\\
J033229.88-274424.4$^{KS}$ & 53.1244949 & -27.7401248 & 16.45 & 0.03 & 2.23 &1.09 & Clean & Disk galaxy &...& ...\\
J033229.98-274529.9$^{KS}$ & 53.1249148 & -27.7583013 & 21.06 & 0.93 & 4.38 &
2.62&Clean & Core + extended emission &1.2091$^{S}$& -23.55\\
J033229.99-274404.8&53.1249588 & -27.7346753 & 16.84 & 0.03 & 2.32 & 0.84 &Clean
& Disk Galaxy&0.0746$^{S}$ & -20.81\\
J033230.06-274523.5$^{KS}$ & 53.1252547 &-27.756535 & 21.81 & 0.25 & 2.47 & 3.20 & Clean & Resolved &... & ...\\
J033230.22-274504.6$^{KS}$ & 53.1258995 & -27.7512749 & 21.50 & 0.92 & 5.39 &
4.07 &Clean & Core + extended emission &...& ...\\
J033230.36-275133.2 &53.1264805 & -27.8592312 & 25.01 & 0.03 & 2.09 & 22.48 & Normal & Resolved &... & ...\\
J033231.80-275110.4 & 53.1324925 & -27.8528853 & 21.03 & 0.98 & 6.58 & 2.99 & Clean & Unresolved &... & ...\\
J033231.82-275110.6 & 53.1325905 & -27.8529435 & 21.06 & 0.99 & 5.74 & 2.50 &
Clean & Star\tablenotemark{j} &... & ...\\
J033231.94-274531.3 & 53.1330882& -27.7587052 & 21.11 & 0.99 & 2.15 & 1.16 &
Normal & Star\tablenotemark{g} &... & ...\\
J033232.04-274523.9 &53.1334996 & -27.7566329 & 23.45 & 0.03 & 1.97 & 6.95 & Normal & Interacting system &... & ...\\
J033232.12-275636.8 & 53.1338346 & -27.9435593 & 21.04 & 0.99 & 1.97 & 1.02 &
Normal & Star\tablenotemark{b} &... & ...\\
J033232.32-274316.4 & 53.1346778& -27.7212189 & 22.11 & 0.98 & 2.24 & 2.16 & Clean & Unresolved &... & ...\\
J033232.49-275044.0 & 53.1353687 & -27.8455431 & 23.42 & 0.03 & 2.12 & 8.15 & Normal & Disk galaxy &... & ...\\
J033232.61-275316.7 & 53.1358826 & -27.8879636 & 22.93 & 0.03 & 2.32 &5.00&Clean & Interacting system &0.9873$^{S}$& -21.14\\
J033232.67-274944.6&53.1361117 & -27.829048 & 18.29 & 0.03 & 2.95 & 0.83 & Clean & Elliptical &... & ...\\
J033233.68-274035.6 & 53.1403383 & -27.6765489 & 24.94 & 0.02 & 2.59 & 24.37 & Clean & Resolved &... & ...\\
J033235.38-274704.3 & 53.1474321 & -27.7845155 & 25.21 & 0.04 & 1.97 & 20.68 & Normal & Resolved &... & ...\\
J033236.92-275308.4 & 53.1538193 & -27.885679 & 24.41 & 0.03 & 1.95 & 11.75 & Normal & Resolved &... & ...\\
J033237.16-274128.2 & 53.1548456 & -27.6911768 & 23.36 & 0.03 & 2.49 & 7.93 & Clean & Resolved &... & ...\\
J033237.93-274609.1$^{C}$ & 53.1580245 & -27.769192 & 19.96 & 0.93 & 4.37 &
1.51& Clean & Unresolved & 0.086\tablenotemark{k} & -18.02\\
J033238.12-273944.8$^{KS}$ & 53.1588307 & -27.6624444 & 20.43 & 0.92 & 13.09 & 5.39&Clean & Core + extemded emission &0.8376$^{S}$& -23.20\\
J033238.89-275406.7&53.1620545 & -27.9018501 & 24.49 & 0.03 & 1.99 & 18.00 & Normal & Resolved &... & ...\\
J033239.09-274601.8$^{C,KS}$ & 53.1628593 & -27.7671602 & 20.96 & 0.94 & 7.88 &
4.39 &Clean & Core + extended emission &... & ...\\
J033239.47-275300.5 & 53.1644567 & -27.8834689 & 20.54 & 0.03 & 2.01 & 1.23 &Normal & Elliptical &...& ...\\
%J033239.97-274213.6 & 53.1665349 & -27.7037912 & 22.36 & 0.03 & 4.46 & 9.07 &
%Clean & Diffractio &... & ... & ...\\
J033240.27-274949.7 & 53.1678017 &-27.830481 & 24.69 & 0.02 & 1.99 & 17.90 & Normal & Complex &None & ...\\
J033240.89-275449.2 & 53.1703678 & -27.9136643 & 25.14 & 0.02 & 2.01 & 17.08 & Normal & Resolved &... & ...\\
J033241.05-275234.1 & 53.1710547 & -27.8761468 & 20.72 & 0.97 & 2.71 & 1.08
&Clean & Star\tablenotemark{e} &0.0000$^{S}$& ...\\
J033241.87-274651.1 &53.1744478 & -27.7808655& 23.35 & 0.01 & 2.01 & 12.07 & Normal & Tadpole
galaxy\tablenotemark{h} &... & ...\\
J033242.61-275453.8& 53.1775348 & -27.9149453 & 20.82 & 0.99 & 2.64 & 1.25&Clean & Star\tablenotemark{e} &0.0000$^{S}$& ...\\
J033243.24-274914.2$^{KS,T}$&53.1801493 & -27.8206046 & 22.49 & 0.93 & 9.56 &
12.41&Clean & Core + extended emission &0.2145$^{T}$& -17.66\\
J033243.93-274351.1 &53.1830401 & -27.7308524 & 24.26 & 0.04 & 2.00 & 9.75 & Normal & Resolved &... & ...\\
J033244.10-275212.9 & 53.1837535 & -27.8702564 & 20.88 & 0.94 & 2.33 & 0.90 &
Clean & Star\tablenotemark{j} &... & ...\\
J033245.02-275207.7 & 53.1875887& -27.8688155 & 22.57 & 0.03 & 4.45 & 7.33 & Clean & Disk galaxy &... & ...\\
J033246.37-274912.8 & 53.1932061 & -27.8202112 & 21.94 & 0.03 & 2.08 & 2.07&Normal & Elliptical &0.6830$^{S}$ & -21.15\\ J033246.39-274820.1 & 53.1932909
&-27.8055737 & 21.16 & 0.97 & 2.59 & 1.45 & Clean & Star\tablenotemark{i} &... & ...\\
J033247.53-275159.9 & 53.1980298 & -27.8666421 & 21.01 & 0.99 & 2.11 & 0.95 & Normal & Star &... & ...\\
J033247.98-274855.7 & 53.1999289 & -27.8154702 & 20.56 & 0.04 & 2.61 & 1.33&Clean & Elliptical &0.2340$^{S}$ & -19.79\\
J033251.22-275418.3 & 53.2134115&-27.905076 & 24.25 & 0.03 & 1.95 & 13.30 & Normal & Resolved &... & ...\\
J033252.88-275119.8$^{KS}$ & 53.2203537 & -27.8555099 & 21.84 & 0.12 & 2.93&4.23&Clean & Disk galaxy &1.2283$^{S}$ & -22.81\\
J033253.44-275001.4 & 53.2226606 &-27.8337103 & 24.44 & 0.00 & 1.96 & 15.49 & Normal & Complex &...& ...\\
J123553.12+621037.5 & 188.9713311 & 62.1770907 & 21.26 & 0.93 & 5.71 & 4.78 & Clean & Interacting system &... & ...\\
J123556.88+621117.3 & 188.9870032 & 62.1881435 & 20.23 & 0.92 & 6.41 & 2.34 & Clean & Unresolved &... & ...\\
J123557.62+621024.7 & 188.9900912 & 62.1735292 & 22.62 & 0.94 & 4.55 & 6.53 &Clean & Core + extended emission &... & ...\\
J123603.82+621039.3 &189.0159143 & 62.1775909 & 24.48 & 0.00 & 2.60 & 21.88 & Clean & Interacting system &... & ...\\
J123605.75+621356.1 & 189.0239559 & 62.2322631 & 20.82 & 0.99 & 4.01 & 1.93 & Clean & Unresolved &... & ...\\
J123606.47+621506.4 & 189.0269558 & 62.2517698 & 21.42 & 0.89 & 1.98 & 1.28 &Normal & Core + extended emission &... & ...\\
J123612.61+621238.4 &189.0525622 & 62.2106579 & 20.18 & 0.89 & 3.11 & 1.05 &Clean & Core + extended emission &... & ...\\
J123617.99+621635.3 &189.0749736 & 62.2764767 & 20.18 & 0.87 & 9.25 & 4.21 &Clean & Core + extended emission &... & ...\\
J123619.57+620715.2 &189.0815546 & 62.1208771 & 25.14 & 0.02 & 2.04 & 22.65 & Normal & Resolved &... & ...\\
J123621.26+621640.4 & 189.0885649 & 62.2778893 & 21.57 & 0.03 & 1.94 & 1.88 & Normal & Interacting system &... & ...\\
J123622.94+621527.0 & 189.095583 & 62.2574929 & 20.27 & 0.90 & 4.52 & 1.70 & Clean & Core + extended emission &... & ...\\
J123627.48+621406.4 &189.1144932 & 62.2351164 & 25.22 & 0.04 & 2.25 & 25.05 & Clean & Resolved &... & ...\\
J123627.98+621508.1 & 189.1165918 & 62.2522625 & 21.86 & 0.03 & 1.94 & 2.33 & Normal & Interacting system &... & ...\\
J123629.44+621513.3 & 189.1226602 & 62.2536973 & 23.70 & 0.98 & 1.97 & 5.83 & Normal & Unresolved &... & ...\\
J123629.68+621734.7 & 189.1236482 & 62.2929798 & 22.45 & 0.98 & 2.27 & 2.73 & Clean & Unresolved &... & ...\\
J123631.68+620848.7 & 189.1320148 & 62.1468517 & 22.56 & 0.03 & 2.14 & 3.24 & Normal & Elliptical &... & ...\\
J123631.70+620752.3 & 189.1320712 & 62.1311963 & 24.12 & 0.03 & 2.19 & 9.17 & Clean & Resolved &... & ...\\
J123632.50+620701.7 & 189.1354166 & 62.1171419 & 25.37 & 0.82 & 2.03 & 20.07 & Normal & Resolved &... & ...\\
J123633.23+620834.9 & 189.1384756 & 62.1430369 & 21.08 & 0.03 & 5.24 & 3.87 & Clean & Interacting system &... & ...\\
J123636.51+620806.4 & 189.1521369 & 62.1350985 & 22.29 & 0.03 & 2.06 & 3.24 & Normal & Elliptical &... & ...\\
J123637.85+620724.1 & 189.1576928 & 62.1233725 & 22.21 & 0.00 & 1.94 & 5.06 & Normal & Complex &... & ...\\
J123641.44+620730.4 & 189.1726778 & 62.1251072 & 22.62 & 0.00 & 1.94 & 5.58 &Normal & Disk galaxy &... & ...\\
J123648.32+621250.1 & 189.2013266 & 62.2139299 & 20.39 & 0.93 & 2.93 & 1.02 & Clean & Core + extended emission &...& ...\\
J123650.44+620749.7 & 189.2101509 & 62.1304753 & 23.62 & 0.03 & 1.98 & 6.34 & Normal & Resolved &... & ...\\
J123650.75+621439.9 & 189.2114589 & 62.2444226 & 23.73 & 0.03 & 2.13 & 9.86 &Normal & Disk galaxy &... & ...\\
J123651.32+621751.2 & 189.2138382 &62.2975491 & 24.46 & 0.03 & 1.98 & 13.91 & Normal & Resolved &... & ...\\
J123652.44+620959.9 & 189.2185084 & 62.1666296 & 22.44 & 0.98 & 2.23 & 2.58 & Clean & Unresolved &... & ...\\
J123654.99+621635.1 & 189.2291393 & 62.2764217 & 25.46 & 0.97 & 2.31 & 22.23 & Clean & Unresolved &... & ...\\
J123655.90+620828.3 & 189.2329249 & 62.1411947 & 21.01 & 0.99 & 2.85 & 1.49 & Clean & Unresolved &... & ...\\
J123656.51+620837.7 & 189.2354508 & 62.1438079 & 20.75 & 0.99 & 3.38 & 1.56 & Clean & Unresolved &... & ...\\
J123656.91+621950.3 & 189.2371093 & 62.330639 & 21.61 & 0.03 & 2.07 & 2.67 &Normal & Disk galaxy &... & ...\\
J123700.71+621854.4 & 189.2529566 & 62.3151155 & 23.58 & 0.03 & 2.20 & 9.72 &Clean & Interacting system &... & ...\\
J123700.88+621129.5 & 189.2536832& 62.1915322 & 20.32 & 0.97 & 3.40 & 1.22 &
Clean & Star\tablenotemark{d} &... & ...\\
J123701.55+622103.9 & 189.2564508& 62.3510871 & 23.02 & 0.06 & 2.04 & 3.85 & Normal & Elliptical &... & ...\\
J123702.09+621737.8 & 189.2586984 & 62.2938331 & 20.43 & 0.99 & 2.02 & 0.79 & Normal & Unresolved &... & ...\\
J123702.72+621543.9 & 189.2613473 & 62.2621972 & 19.41 & 0.03 & 2.18 & 0.91 &Clean & Disk galaxy &... & ...\\
J123704.80+621455.2 & 189.2699968 & 62.2486665 & 20.99 & 0.98 & 2.79 & 1.44 & Clean & Unresolved &... & ...\\
J123705.48+621526.8 & 189.2728214 & 62.257434 & 24.62 & 0.03 & 2.16 & 14.31 & Normal & Resolved &... & ...\\
J123706.25+622136.9 & 189.2760529 & 62.3602611 & 20.90 & 0.90 & 10.05 & 8.36 & Clean & Disk galaxy &... & ...\\
J123706.87+621702.5 & 189.2786271 & 62.2840158 & 19.86 & 0.92 & 3.81 & 1.27
&Clean & Core + extended emission
&1.0200\tablenotemark{c} & -24.29\\
J123706.93+621429.9 & 189.2788629 & 62.2416441 & 20.70 & 0.98 & 2.77 & 1.24 & Clean & Unresolved &... & ...\\
J123707.49+622148.1 & 189.2812076 & 62.3633707 & 22.03 & 0.92 & 3.75 & 3.83 & Clean & Elliptical &... & ...\\
J123708.35+621105.8 & 189.284812 & 62.184941 & 23.47 & 0.03 & 2.00 & 6.04 & Normal & Resolved &... & ...\\
J123708.65+621051.5 & 189.286038 & 62.1809766 & 20.67 & 0.03 & 2.03 & 1.91 &Normal & Interacting system &...& ...\\
J123716.68+621733.6 &189.3194798 & 62.2926732 & 22.12 & 0.72 & 2.97 & 3.74 & Clean & Elliptical &...& ...\\
J123717.79+622034.2 & 189.3241356 & 62.3428338 & 23.91 & 0.00 & 1.94 & 26.94 & Normal & Resolved &...& ...\\
J123717.82+621130.5 & 189.3242369 & 62.1918003 & 24.28 & 0.17 & 2.19 & 10.70 & Clean & Resolved &...& ...\\
J123719.47+621320.4 & 189.3311376 & 62.2223455 & 20.79 & 0.96 & 2.56 & 1.20 &Clean & Core + extended emission &...& ...\\
J123720.16+621518.9 &189.3339866 & 62.2552618 & 20.16 & 0.88 & 6.72 & 2.39 &Clean & Core + extended emission &...& ...\\
J123723.70+621200.3 &189.3487632 & 62.2000936 & 24.82 & 0.00 & 1.97 & 21.53 & Normal & Resolved &... & ...\\
J123723.72+622113.3 & 189.3488236 & 62.3537049 & 23.62 & 0.98 & 2.30 & 6.61 & Clean & Unresolved &...& ...\\
J123724.77+622103.0 & 189.3531954 & 62.3508303 & 23.80 & 0.00 & 1.96 & 14.08 & Normal & Resolved &...& ...\\
J123728.43+622044.8 & 189.3684533 & 62.3457887 & 22.69 & 0.02 & 2.24 & 7.50 & Clean & Resolved &...& ...\\
J123728.95+621127.8 & 189.3706346 & 62.1910553 & 21.75 & 0.03 & 6.06 & 6.20 &
Clean & Supernova\tablenotemark{f} &...& ...\\
J123729.58+621557.8 & 189.3732355& 62.2660663 & 19.91 & 0.85 & 2.12 & 0.60 &Normal & Core + extended emission &...& ...\\
J123732.41+621751.4 &189.3850213 & 62.2976196 & 24.58 & 0.03 & 2.55 & 17.65 & Clean & Resolved &...& ...\\
J123736.11+621619.1 & 189.4004397 & 62.2719614 & 24.07 & 0.03 & 2.91 & 15.16 & Clean & Resolved &...& ...\\
J123736.59+621632.7 & 189.4024445 & 62.2757619 & 25.03 & 0.03 & 3.09 & 37.46 & Clean & Resolved &...& ...\\
J123738.83+622024.0 & 189.4117843 & 62.3399951 & 20.20 & 0.96 & 3.41 & 1.24 &Clean & Core + extended emission &...& ...\\
J123740.64+622007.9 &189.4193208 & 62.3355268 & 22.60 & 0.03 & 2.50 & 5.41 &Clean & Disk galaxy &...& ...\\
J123741.21+621925.2 & 189.4217126 & 62.3236767 & 21.04 & 0.98 & 2.84 & 1.50 & Clean & Unresolved &...& ...\\
J123741.38+621540.2 & 189.4224198 & 62.2611689 & 25.19 & 0.02 & 2.11 & 20.13 & Normal & Resolved &...& ...\\
J123742.12+621903.0 & 189.425516 & 62.3174941 & 24.48 & 0.04 & 3.43 & 29.44 & Clean & Resolved &...& ...\\
J123742.53+621812.2 & 189.4272043 & 62.3033909 & 21.31 & 0.98 & 5.40 & 3.36 & Clean & Unresolved &...& ...\\
J123746.85+621624.2 & 189.4452252 & 62.2733875 & 21.94 & 0.98 & 2.27 & 1.96 & Clean & Unresolved &...& ...\\
J123749.58+621346.6 & 189.4565982 & 62.2296055 & 24.56 & 0.98 & 2.49 & 14.73 & Clean & Unresolved &...& ...\\
J123754.25+621853.0 & 189.47606 & 62.3147251 & 24.84 & 0.02 & 2.21 & 20.84 & Clean & Resolved &...& ...\\
\enddata
\tablenotetext{a}{\cite{groenewegen_eso_2002}}
\tablenotetext{b}{\cite{hatziminaoglou_eso_2002}}
\tablenotetext{c}{\cite{hornschemeier_chandra_2001}}
\tablenotetext{d}{\cite{mendez_starcounts_1998}}
\tablenotetext{e}{\cite{popesso_great_2009}}
\tablenotetext{f}{\cite{riess_type_2004}}
\tablenotetext{g}{\cite{santini_star_2009}}
\tablenotetext{h}{\cite{straughn_tracing_2006}}
\tablenotetext{i}{\cite{taylor_public_2009}}
\tablenotetext{j}{\cite{wolf_calibration_2008}}
\tablenotetext{k}{\cite{szokoly_chandra_2004}}
\end{deluxetable}

\end{document}